\documentclass[aps,prb,twocolumn,superscriptaddress,citeautoscript,nobibnotes,nofootinbib,10pt]{revtex4-2}
\usepackage{graphicx}
\usepackage{microtype}
\usepackage{lmodern}
\usepackage{graphicx}
\usepackage[
    format=plain,
    figurename=Figure
]{caption}

\usepackage{ragged2e}
\DeclareCaptionJustification{plain}{\justifying}
\usepackage[subrefformat=simple,labelformat=simple]{subcaption}

\usepackage{amsmath}
\usepackage{amssymb}
\usepackage{mathrsfs}
\usepackage{physics}

\usepackage{siunitx}
\AtBeginDocument{\RenewCommandCopy\qty\SI}
\DeclareSIUnit\hopping{\mathit t_0}

\usepackage{hyperref}
\hypersetup{
    colorlinks=true,
    urlcolor=blue,
    citecolor=blue,
    linkcolor=blue
}
\usepackage[capitalize]{cleveref}
\crefname{figure}{Fig.}{Figures}
\crefname{section}{Sec.}{Sections}

\usepackage{csquotes}

\DeclareMathOperator*{\mean}{mean}

\usepackage{soul}
\newcommand{\changes}[1]{{#1}}
\newcommand{\vanishchanges}[1]{}

\begin{document}
\title{Hybrid quantum--classical matrix-product state and Lanczos methods for electron--phonon systems
with strong electronic correlations: Application to disordered systems coupled to Einstein phonons}

\author{Heiko Georg Menzler}%
\thanks{These authors contributed equally.}%
\affiliation{Institut f\"{u}r Theoretische Physik, Georg-August-Universit\"{a}t G\"{o}ttingen, D-37077 G\"{o}ttingen, Germany}
\author{Suman Mondal}%
\thanks{These authors contributed equally.}%
\affiliation{Max Planck Institute for the Physics of Complex Systems, N\"{o}thnitzer Str. 38, 01187 Dresden, Germany}
\author{Fabian Heidrich-Meisner}%
\affiliation{Institut f\"{u}r Theoretische Physik, Georg-August-Universit\"{a}t G\"{o}ttingen, D-37077 G\"{o}ttingen, Germany}

\date{\today}

\begin{abstract}
We present two quantum--classical hybrid methods for simulating the time-dependence of electron--phonon systems that treat electronic correlations numerically exactly and optical-phonon degrees of freedom classically. These are a time-dependent Lanczos and a matrix-product state method, each combined with the multi-trajectory Ehrenfest approach. Due to the approximations, reliable results are expected for the adiabatic regime of small phonon frequencies. We discuss the convergence properties of both methods for a system of interacting spinless fermions in one dimension and provide a benchmark for the Holstein chain. As a first application, we study the decay of charge density wave order in a system of interacting spinless fermions coupled to Einstein oscillators and in the presence of quenched disorder. We investigate the dependence of the relaxation dynamics on the electron--phonon coupling strength and provide numerical evidence that the coupling of strongly disordered systems to classical oscillators leads to delocalization, thus destabilizing the (finite-size) many-body localization in this system.
\end{abstract}

\maketitle

\section{Introduction} 
Electron--phonon coupling is ubiquitous in many phenomena
in solid-state physics and quantum chemistry.
In solid-state physics, the importance of phonons for electronic properties is evident in transport measurements, optical excitations, for the formation of Cooper pairs in conventional BCS superconductors~\cite{BCS1957}, and in the physics of polarons~\cite{Frohlich1954}.
More recently, the role of phonons as a relaxation channel for electronic excitations became experimentally accessible due to the development of ultrafast dynamics techniques~\cite{Giannetti2016} and is also discussed in the context of energy-conversion processes~\cite{Pastor2022}.
Subsequently, with similar tools, the possibility of modifying electronic properties via optically targeting specific phonon modes has been explored (see~\cite{Bloch2022} for a review and~\cite{Rini2007,Tobey2008} for examples). 
In quantum chemistry, the role of vibrations is obvious for the excitation spectra of molecules and for certain reaction processes, to name a few examples~\cite{Nibbering2005}.

The theoretical treatment of electron--phonon problems becomes particularly challenging when additional interactions or degrees of freedom need to be accounted for. For instance, one can think of materials with orbital, spin, and phonon degrees of freedom with a substantial mutual coupling~\cite{Dagotto2001,Tokura2006}. In addition, whenever bare or induced electronic interactions push the system out of the validity range of perturbation theory, mean-field or density-functional theory approaches, then, there is a limited set of unbiased many-body techniques.

Consequently, there has been recent activity in combining computational many-body methods that can capture strong electronic correlations with an efficient treatment of the phonon sector.
These include (quantum)-Monte-Carlo methods~\cite{Weber2021,Cohen-Stead2023}, dynamical mean-field theory~\cite{Picano2023}, hierarchical equations of motions~\cite{Jankovic2022}, Lanczos techniques~\cite{Bonca1999,Dorfner2015}, Green's function methods~\cite{Goodvin2008}, matrix-product state approaches~\cite{Jeckelmann1998,Guo2011,Brockt2015,Koehler2021,Jansen2020,Stolpp2021,Jansen2022,Jansen2023}.

In our work, we are concerned with wave-function methods in finite-dimensional Hilbert spaces such as Lanczos~\cite{Manmana2005} or matrix-product state (MPS) methods~\cite{Schollwoeck2011}.
In previous MPS and density-matrix renormalization group developments, several ways of dealing with quantum phonons were   introduced, including the pseudosite technique~\cite{Jeckelmann1998}, local basis optimization~\cite{Zhang1998}, and projected purification~\cite{Koehler2021}.
However, the regime of slow phonons with phonon energies much smaller than the bandwidth remains hard to access for these approaches~\cite{Stolpp2021}.
In typical solid-state settings, however, this adiabatic regime is often the physically relevant one~\cite{Giustino2017}.

As a compromise, one can resort to well-established semi-classical methods that are widely used in connection with \textit{ab initio} approaches~\cite{Lively2023} or in quantum chemistry.
The simplest such technique is the multi-trajectory Ehrenfest method (MTE)~\cite{Aleksandrov1981,Tully1998,Nielsen2001}, which treats the phonon dynamics classically.
The quantum mechanical uncertainty is accounted for by a sampling of the initial coordinates and momenta from the initial quantum state probability distributions (often assumed to be the ground states), resulting in independent trajectories. 
We present the successful implementation of the MTE approach with a full quantum mechanical treatment of strongly correlated electrons in both time-dependent Lanczos and MPS simulations.
This complements the usage of MTE with electrons that are only weakly electronically correlated and where such correlations are \changes{accounted for} with approximations such as Hartree-Fock (see, e.g.,~\cite{Li2005}) or density functional theory~\cite{Lively2023}.

MTE is formally equivalent to the truncated Wigner approximation~\cite{Polkovnikov2010} of the phonon sector combined with a mean-field treatment of the electron--phonon coupling.
In an earlier work, it was shown that MTE for the one-dimensional Holstein model yields numerically accurate results for electronic observables in the adiabatic regime for certain initial conditions~\cite{TenBrink2022}.
In that case, the actual problem solved with MTE consists of non-interacting electrons whose wave function depends on the time-dependent momenta and coordinates of the oscillators.
Thus, essentially, single-electron orbitals are propagated.
Here, we extend this to a many-body problem of electrons that interact already in the absence of an electron--phonon coupling and hence a propagation of the fully many-body wave function is required.
This can be accomplished in one spatial dimension.

A related algorithm has recently been presented by Manawadu et al.~\cite{Manawadu2022,Manawadu2023},
where a time-dependent DMRG algorithm is combined with Ehrenfest dynamics for phonons, in a single-trajectory formulation, based on the original block-version of the density matrix renormalization group method~\cite{White1992,White1993}. 
Our implementation is based on MPS and therefore, can easily be adapted to new algorithmic developments that are largely formulated in the MPS language.
In particular, one can incorporate modern time-evolution schemes such as the time-dependent variational principle~\cite{Paeckel2019}.

We note that similar ideas were recently pursued in Monte-Carlo simulations~\cite{Weber2021} and in dynamical mean-field theory~\cite{Picano2023}, yet both lack an exact incorporation of electronic correlations.
There are also many extensions to MTE that account for its known shortcomings such as surface-hopping methods~\cite{Subotnik2016} or coupled-trajectory methods such as the multi-configurational Ehrenfest approach~\cite{Shalashilin2009,Shalashilin2010}.
While we consider it worth to pursue such extensions in the future, we here aim at establishing the simplest and computationally inexpensive approach first.

As an application, we consider the temporal decay of charge density wave (CDW) order. Such CDW order can either be present in the equilibrium phase diagram of interacting electrons
and may decay due to external perturbations or we may envisage the initial CDW as being a result of some prior perturbations, e.g., in a pump-and-probe experiment (see, e.g.,~\cite{Huang2024}).
For the case of interest here, i.e., one-dimensional systems, the decay of CDW order of spinless fermions due electronic interactions and/or electron--phonon interactions has been studied previously~\cite{Hashimoto2017,Stolpp2020}.

Moreover, we are specifically interested in the decay of CDW order in systems with quenched and uncorrelated disorder as this relates to the stability of a putative many-body localization (MBL) in a system of interacting electrons~\cite{Nandkishore2015,Abanin2019}. We will show clear evidence that the coupling to classical phonons in the adiabatic regime leads to a decay of CDW order, indicating the instability of MBL in such a system.
Furthermore, we will study the spreading of a single polaron in a disorder potential (see~\cite{Bronold2002,Bronold2004,Das2008,Berciu2010,Berciu2012,Tozer2014,Kloss2019}) which suggests sub-diffusive dynamics, consistent with~\cite{Prelovsek2018}. Our analysis of the CDW order decay hints at the presence of subdiffusive dynamics even at a finite filling, at least at intermediate times.

The broader topic of thermalization in electron--phonon systems, studied from a model Hamiltonian perspective, 
has, of course, a long history, beyond what can reasonably be reviewed in this article.
Recently, for instance, the dynamics after quenches in the electron--phonon coupling were studied using DMFT~\cite{Murakami2015}.
The \changes{combination of disorder} and electron--phonon coupled system, already in the absence of \changes{direct electronic interactions}, leads to partial delocalization as evidenced in, e.g.,~\cite{DiSante2017}, related to bad metal physics and variable-range hopping~\cite{Fleishman1978}.
Note that the stability of many-body localization in the presence of a variety of  different baths has been studied previously~\cite{HuseNandkishore2015,Sarang2017,Bar2022,Abanin2022,SerbynAbanin2022,Sierant2023b}.

The plan of the paper is the following. In \cref{sec:model}, we will introduce a model of interacting spinless fermions in the presence of disorder and electron--phonon coupling of the Holstein type thus combining the $t$--$V$ model (equivalent to spin-1/2 Heisenberg chains) with the Holstein model.
In \cref{sec:methods}, we briefly review the MTE technique and then describe its incorporation into time-dependent Lanczos and time-evolving block decimation (TEBD) approaches.
We will discuss convergence properties of both these many-body techniques, the Lanczos-MTE and the TEBD-MTE. In \cref{sec:anderson}, we investigate the stability of an Anderson insulator against a coupling to classical and adiabatic phonons, both for a single electron and for finite electronic densities, while in \cref{sec:mbl}, we turn to the interacting case that is connected to MBL.
We present a summary and an outlook in \cref{sec:conclusions}.

\section{Model}
\label{sec:model}
We consider spinless fermions in a one-dimensional lattice of $L$ sites with open boundary conditions, where $\hat c_{\ell}$ is the annihilation operator of a fermion on site $\ell$. 
The Hamiltonian contains five parts:
\begin{align}
    \label{eq:hamiltonian}
    \hat H &= \hat H_\text{kin} + \hat H_\text{in}+ \hat H_\text{e-ph} + \hat H_\text{ph} + \hat H_\text{dis} \,,
\end{align}
the electronic kinetic energy (i.e., hopping term on a lattice), the electronic interaction, the electron--phonon coupling term, quantum mechanical Einstein phonons, and the disorder term, respectively.
These terms are individually given by:
\begin{eqnarray}
 \hat H_{\text{kin}} &=& -\unit{\hopping} \sum_{\ell} (\hat c^\dagger_{\ell+1} \hat c_{\ell} + \mathrm{h.c.})\,,\\
 \hat H_{\text{in}} &=& V \sum_{\ell} \hat n_{\ell} \hat{n}_{\ell+1} \,,\\
 \hat H_\text{e-ph} &=& \gamma \sum_{\ell} \hat n_{\ell} (\hat b^\dagger_\ell +\hat b_{\ell}) \,,\\
 \hat H_\text{ph} &=& \omega_0 \sum_{\ell} \hat b_{\ell}^\dagger \hat b_{\ell}\,,\\
  \hat H_\text{dis} &=& \sum_\ell \epsilon_\ell \hat n_{\ell} \,.
\end{eqnarray}
When only the electronic hopping-matrix element $\unit{\hopping}$ and the electronic nearest-neighbor interaction $V$ are nonzero, we obtain the $t$--$V$ model that is equivalent to the spin-1/2 XXZ chain, with $V=\qty{2}{\hopping}$ corresponding to the SU(2) symmetric Heisenberg chain. 
Here, $\hat n_\ell=\hat c^\dagger_{\ell} \hat c_{\ell} $ is the electronic occupation operator at site $\ell$, $\omega_0$ is the frequency of Einstein oscillators, $\hat b_{\ell}$ creates a phonon on site $\ell$ and $\gamma$ denotes the electron--phonon coupling strength. 
Therefore, when only $\unit{\hopping}\not=0$, $\omega_0 \not=0$ and $\gamma\not=0$, then we obtain the Holstein chain. 
Note that $\hat X_\ell = \frac{1}{\sqrt{2}}(\hat b^\dagger_\ell +\hat b_{\ell})$ is the displacement operator of a local oscillator. 
We draw the onsite potentials $\epsilon_\ell \in [-W,W]$ from a box distribution of width $W$. 

As the initial state, throughout this work, we consider an ideal charge density wave state with one particle on every other site (assuming $L$ even):
\begin{align}
\label{eq:cdw_state}
    |\psi(t=0) \rangle =
    \prod_{\ell=1}^{L/2} \hat c_{2\ell-1}^\dagger |0\rangle\,,
\end{align}
where $|0\rangle$ is the vacuum state.
The associated order parameter is $\mathcal{O}_\mathrm{CDW}$ with
\begin{align}
    \mathcal{O}_\mathrm{CDW}
    = \frac{1}{N} \sum\limits_{\ell=1}^L (-1)^{\ell{+1}} \langle \hat{n}_\ell \rangle
    \,,
\end{align}
with the total number of particles in the system $N$.
We will also consider the spreading of a single particle initially localized at site $\ell_0$ coupled to classical phonons using the initial state:
\begin{align}
    |\psi(t=0) \rangle =
    \hat c_{\ell_0}^\dagger | 0 \rangle \,.
\end{align}

In the context of MBL, one often studies the so-called imbalance $I$~\cite{Schreiber2015}, defined as
\begin{align}
I_\mathrm{CDW} & = {\frac{N_{\text{odd}}-N_{\text{even}}}{N_{\text{odd}}+N_{\text{even}}}}\,,\\
{N_{\text{odd}}} & {= \sum_{\ell=1}^{L/2} \langle \hat n_{2\ell-1} \rangle \,,}\\
{N_{\text{even}} }& {= \sum_{\ell=1}^{L/2} \langle \hat n_{2\ell} \rangle\,.}
\end{align}
$I_\mathrm{CDW}$ is identical to the charge-density wave order parameter $\mathcal{O}_\mathrm{CDW}$ for our specific initial state.

\begin{figure}
\centering
    \includegraphics[clip, trim={{0.85\linewidth} {0\linewidth} {0.0\linewidth} {1.\linewidth}}, width=1.\linewidth]{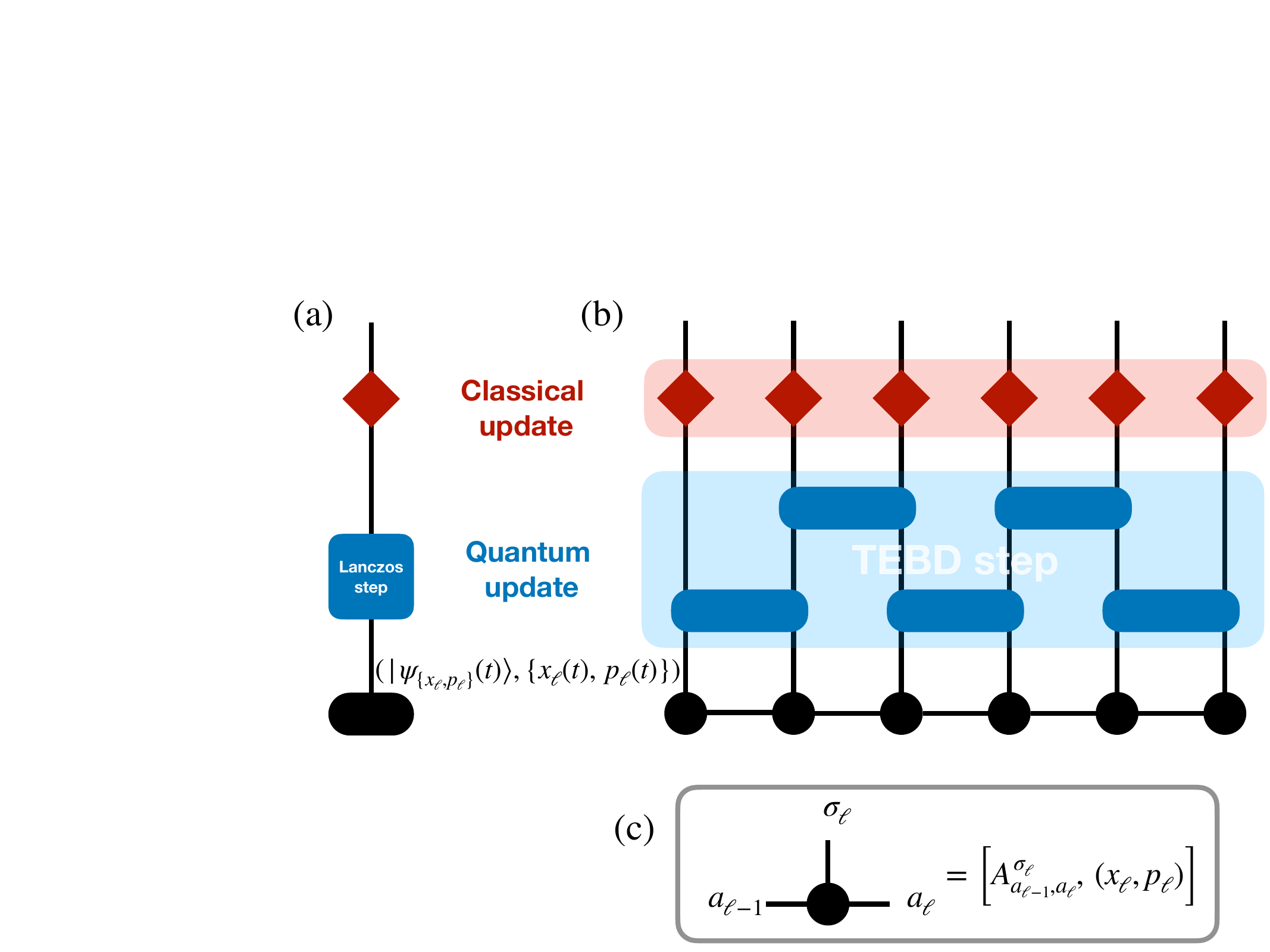}
    {\phantomsubcaption\label{fig:method_lanczos_mte}}
    {\phantomsubcaption\label{fig:method_tebd_mte}}
    {\phantomsubcaption\label{fig:method_mte_tensor}}
    
    \caption{
        \subref{fig:method_lanczos_mte} The Lanczos-MTE and \subref{fig:method_tebd_mte} TEBD-MTE methods are graphically represented for a single time step $\Delta t$.
        \subref{fig:method_lanczos_mte} The lowest layer (elongated black circle) represents the state of quantum and classical sub-system combined at time $t$.
        The evolution of the quantum sub-system using the Lanczos method is represented by the second layer (blue square), and the third layer (red diamond) represents the consecutive evolution for the classical sub-system.
        Similarly, in \subref{fig:method_tebd_mte}, the bottom layer stands for the MPS representation of the quantum sub-system.
        The second layer represents the TEBD step for the quantum sub-system and the last layer evolves the classical-sub system.
        \subref{fig:method_mte_tensor} Element of the bottom layer of \subref{fig:method_tebd_mte} on site $\ell$.
        This contains the information about the MPS in $A^{\sigma_\ell}_{a_{\ell-1,\ell}}$ and the coordinates and momenta for the phonons ($\lbrace x_\ell,~p_\ell \rbrace$) as external parameters.
    }
    \label{fig:method}
\end{figure}

\section{Methods}
\label{sec:methods}
In this section, we provide a brief summary of the MTE method in \cref{sec:mte}. For a detailed discussion, see~\cite{TenBrink2022} for a review. We then describe the incorporation of MTE into time-dependent Lanczos~\cite{Manmana2005} and TEBD~\cite{Vidal2004} in Secs.~\ref{sec:lanczos-mte} and~\ref{sec:DMRG-mte}, respectively. We discuss the convergence of both hybrid methods.

\subsection{Multi-trajectory Ehrenfest dynamics}
\label{sec:mte}

\subsubsection{Basic idea and algorithm}

The multi-trajectory Ehrenfest method~(MTE) can be derived as a special case of the quantum--classical Liouville~(QCL) equation~\cite{Aleksandrov1981,Nielsen2001}, with two essential approximations:
The coupling between the electronic subsystem and the phonon environment is treated in a mean-field approximation and the phonon dynamics is treated classically~\cite{Grunwald2009}.
Intuitively, the latter is a reasonable assumption when the phonon dynamics is much slower then the electronic dynamics.
The QCL equation is based on treating the phonon environment in the phase space formalism of quantum mechanics and can be used to approximate the dynamics of the partially transformed Wigner function~\cite{Wigner1932,Imre1967}.
The full derivation of MTE within this formalism is illuminating, yet for the purpose of our work, we will only need the key algorithmic steps and a discussion of control parameters and convergence.
We refer the reader to~\cite{TenBrink2022} and references therein for a more detailed account and a comparison of MTE to other trajectory-based methods. 

We first discuss the single-trajectory version~\cite{Tully1998} and provide the equations of motion for the Hamiltonian from \cref{eq:hamiltonian} for the quantum evolution of the electronic state $|\psi(t)\rangle$ and the classical evolution of the set of coordinates and momenta $\lbrace x_\ell,p_\ell \rbrace$ of the oscillators. The electronic state depends on the $\lbrace x_\ell,p_\ell \rbrace$, that is,
$|\psi(t)\rangle = |\psi_{\lbrace x_\ell,p_\ell\rbrace }(t)\rangle$.
The classical and quantum Hamiltonians are
\begin{align}
    \label{eq:mte_hamiltonian_env}
    \mathscr{H}_\text{MTE} &= \frac{\omega_0}{2} \sum\limits_\ell x_\ell^2 + p_\ell^2 - \sqrt{2} \gamma \sum\limits_\ell \mel{\psi}{\hat{n}_\ell}{\psi} x_\ell \,,\\
    \label{eq:mte_hamiltonian_subs}
    \hat{H}_\text{MTE} &= \hat{H}_\text{kin} + \hat{H}_\text{in} + \hat{H}_\text{dis} - \sqrt{2} \gamma \sum\limits_\ell \hat{n}_\ell x_\ell
\,.\end{align}
From this, we can directly write down the equations of motion for the classical trajectory $\lbrace x_\ell, p_\ell\rbrace$ and the electronic state:
\begin{align}
    \label{eq:ehrenfest_eom1}
    \frac{d x_\ell}{dt} &= \omega_0 p_\ell\,,\\
    \label{eq:ehrenfest_eom2}
    \frac{d p_\ell}{dt} &= -\omega_0 x_\ell + \sqrt{2} \gamma \mel{\psi(t)}{\hat{n}_\ell}{\psi(t)}\,,\\
    \label{eq:ehrenfest_eom3}
    \pdv{|\psi(t)\rangle}{t} &= -i \hat{H}_\text{MTE} |\psi(t)\rangle
\,.\end{align}
The MTE scheme consist of simulating many single trajectories for a set of given initial conditions $\lbrace x_{\ell}(t=0),p_\ell(t=0)\rbrace$.
The  initial conditions for each trajectory are drawn from the quantum mechanical Wigner function of the initial state in the phonon sector. 
Subsequently, observables are computed for individual trajectories and then averaged over sufficiently many trajectories, with $N_{\text{traj}}$ the number of trajectories. 
Since the equations of motion, by construction, aim to conserve the properties of the Wigner function, we expect that the trajectory of the observable will converge---within the aforementioned approximations---to the quantum operator expectation value with $\sim 1 / \sqrt{N_\mathrm{traj}}$ (see the statistical convergence analysis in~\cite{TenBrink2022}).
Concretely, for our initial state, we sample $\lbrace x_\ell, p_\ell\rbrace$ from a multivariate Gaussian distribution
\begin{align}
    \label{eq:gaussian_distribution}
    P(\{x_\ell, p_\ell\}) \propto \prod_{\ell = 1}^L \exp(-\frac{1}{2}\frac{x_\ell^2 + p_\ell^2}{2})
    \,,
\end{align}
corresponding to the Wigner function of the lowest-energy phonon mode.

The computational effort for systems with disorder, treated later in this manuscript, is considerable because one has to generate multiple trajectories for each disorder realizations.
Indeed, in such cases, we need to trade-off between the desired accuracy and expended computational effort.
We notice that for realizations with strong disorder, the fluctuations in the observables between trajectories of different initial states are small compared to the fluctuations between different disorder realizations.
Therefore we choose to sample each disorder realization only with $N_{\mathrm{traj}}={200}$ trajectories with initial conditions drawn from~\cref{eq:gaussian_distribution} for each of $N_\mathrm{dis}=\SIrange{200}{2000}{}$ different disorder realizations.
This allows us to reduce the absolute error in \changes{$\mathcal{O}_\mathrm{CDW}$, due to insufficient sampling of trajectories}, to $\sim 10^{-2}$, which we estimate by resampling through bootstrapping.

For the data points shown in this manuscript it was necessary to simulate a total of over \qty{10} million individual trajectories.
This is only possible due to the \enquote{embarrassingly parallel} nature of the MTE setup: Every trajectory can be simulated independently which allows for efficient parallel implementations on large computing clusters.
On the other hand, achieving sufficient sampling for systems without disorder is generally multiple orders of magnitude easier than for disordered cases. 

\subsubsection{Time propagation}

The standard way to solve the quantum and classical equations of motion numerically is an alternating time propagation of each subsystem from time $t\to t+\Delta t$ ($\Delta t$ is the time step), with the other subsystem, i.e., $\ket{\psi(t)}$ or $\lbrace x_\ell,p_\ell\rbrace$ kept constant, i.e., a version of a splitting method~\cite{McLachlan2002}. 
This scheme can be viewed as a Trotter-Suzuki breakup of the original full quantum propagator, which, after replacing the Hamiltonian of the phonon sector by its classical counterpart takes the form:
\begin{equation}
\label{eq:splitting_method}
\begin{aligned}
    &e^{-\Delta t \{\mathscr{H}_\text{MTE}, \cdot \} - i \Delta t\hat{H}_\text{MTE}}= \\
    & e^{-\Delta t\{\mathscr{H}_\text{MTE}, \cdot \}} e^{-i \Delta t \hat{H}_\text{MTE}} + \mathcal{O}((\Delta t)^2) \,
\,,\end{aligned}
\end{equation}
where $\{\cdot, \cdot\}$ denotes the Poisson brackets from classical mechanics.
Higher-order versions are possible, which we determine is not necessary for our purpose. Therefore, we expect that the error introduced by the splitting method scales as $(\Delta t)^2$ per time step.

In our actual implementation, we solve the classical equations of motion analytically given the current value of the external force as this is simply a displaced harmonic oscillator for each site $\ell$.
That information is fed back into the quantum propagation. The formal solution for the time propagation of the classical coordinates and momenta from $t $ to $t+\Delta t$ reads:
\begin{align}
\begin{split}
x_\ell(t+\Delta t) &= x_\ell(t) \cos(\omega_0 \Delta t) + p_\ell(t) \sin(\omega_0 \Delta t) \\ 
&\quad+ \frac{\sqrt{2}\gamma}{\omega_0}(1 - \cos(\omega_0 \Delta t))\mel{\psi(t)}{\hat{n}_\ell}{\psi(t)}\,,
\end{split}\\
\begin{split}
p_\ell(t+\Delta t) &= -x_\ell(t) \sin(\omega_0 \Delta t) + p_\ell(t) \cos(\omega_0 \Delta t) \\
& \quad+ \frac{\sqrt{2}\gamma }{\omega_0} \sin(\omega_0 \Delta t)\mel{\psi(t)}{\hat{n}_\ell}{\psi(t)} \,.
\end{split}
\end{align}
In practical simulations, monitoring the conservation of energy yields a criterion for choosing a sufficiently small $\Delta t$.

Obviously, there is a choice in how the quantum subsytem is propagated.
Most applications of MTE treat the electrons as non-interacting or weakly interacting, using Hartree-Fock \cite{Li2005} or ab-initio methods \cite{Lively2023}.
In our work we implement two time-propagation schemes that treat electronic correlations accurately on finite systems, Lanczos and TEBD.

\subsubsection{Discussion of strength and weaknesses of MTE}
\label{sec:strengths_and_weaknesses}
MTE is one of the simplest trajectory based methods~\cite{Ehrenfest1927,Tully1998,Kirrander2020}, stable, and straightforward to implement.
Some limitations can be understood in the Born-Oppenheimer picture. 
For instance, in MTE, electrons experience the average force from several Born-Oppenheimer surfaces and do not see the effects of the individual Born-Oppenheimer surfaces and thus cannot account for splitting of wave packets, etc.

Another known issue in independent trajectory methods is overcoherence, namely the approximations ignore physical
processes that cause decoherence~\cite{Kaeb2002}. Due to the classical nature of the time-evolution in the environment a dephasing or interference of quantum modes can not happen, resulting in overcoherent dynamics~\cite{Kaeb2002}. 
Some of the more sophisticated trajectory-based methods are designed to overcome these shortcomings of MTE, see~\cite{TenBrink2022} and references therein\changes{. H}owever, in many cases, these modifications are based on further assumptions about the system and/or can make the implementation of the algorithm significantly more complex and less scalable.

In the secondary approximation of the MTE method, the mean-field treatment of the electron--phonon coupling, the correlations between subsystem and environment are neglected.
To the best of our knowledge,  no systematic improvement exists to account for the electron--phonon coupling term
beyond the mean-field approximations.
Summarizing, in general, MTE is by no means guaranteed to produce quantitatively correct results and may even lead to unphysical predictions for arbitrary choices of model parameters.

Despite the inherent limitations, nevertheless, for the case of adiabatic phonons $\omega_0 \ll \unit{\hopping}$, there are examples where MTE agrees with exact results. For instance for initial product states in the computational basis, electronic observables exhibit very small errors when compared to full quantum methods, while phonon observables qualitatively follow the correct dynamics~\cite{TenBrink2022}. These observations refer to expectation values only.
The influence of the initial state on the quality of the agreement between MTE and exact methods has recently been discussed by Paprotzki et al.~\cite{Paprotzki2023}. Certain terms in the short-time expansion cancel out
for  initial states that are diagonal in the computational basis,
yet play a role for, e.g., Slater determinants in the electronic quasimomentum eigenbasis. For other initial states with a finite density of electrons, e.g., a Fermi sea, phononic occupations may even become negative at large values of $\omega_0 \sim 2 \unit{\hopping}$~\cite{Paprotzki2023}. This relates to the well-known fact that in the long-time, stationary state, occupations obtained from Ehrenfest dynamics may deviate
from the correct result \cite{Parandekar2005,Parandekar2006}.

Concerning the mere convergence with respect to its actual numerical control parameters, the statistical and numerical errors controlled by $N_\mathrm{traj}$ and $\Delta t$, convergence with respect to these parameters can easily be achieved in MTE.
This makes MTE a technically simple-to-implement and to control technique, rendering it a reasonable starting point for hybrid methods for quantum many-body problems. While beyond the scope of paper, we envision that in the future, Lanczos or TEBD may also be combined with coupled trajectory methods like multi-configurational Ehrenfest~\cite{Shalashilin2009,Shalashilin2010} which at least has internal control parameters that allow convergence to the physically correct results in principle, as was demonstrated in~\cite{TenBrink2022}. 

\subsection{Lanczos-MTE method}
\label{sec:lanczos-mte}

\begin{figure}[t]
    \centering
    \includegraphics[width=\linewidth]{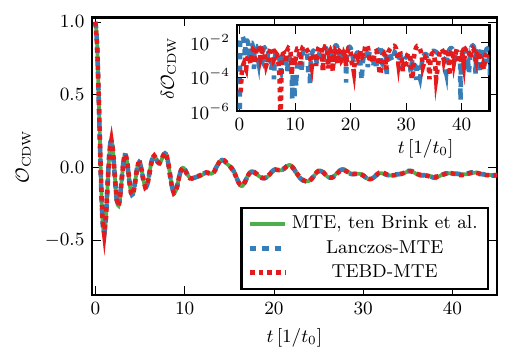}
    \caption{
        Comparison of TEBD-MTE (${\delta = 10^{-8}}$), Lanczos-MTE ($\epsilon = 10^{-9} / t_\text{f}$), and regular MTE as implemented in Ref.~\cite{TenBrink2022} for a non-interacting system ($L=13$, $V=0$, $\omega_0=\qty{0.1}{\hopping}$, $\gamma=\qty{0.4}{\hopping}$, $W=0$).
        Lanczos-MTE and TEBD use $\Delta t = \qty{0.01}{\per\hopping}$ and all methods are averaged over $N_\mathrm{traj}=4000$ trajectories.
        The inset shows the difference $\delta \mathcal{O}_\mathrm{CDW}$ between the many-body hybrid techniques and the reference data from regular MTE from Ref.~\cite{TenBrink2022}.
    }
    \label{fig:mte_comparison_noninteracting}
\end{figure}

In \cref{sec:mte} we have seen how the time-evolution operator of the system can be broken down into independent time-evolution operators of the subsystem and the environment.
Clearly, we have not constrained how either of the time-evolution operators must be implemented.
Therefore, we are free to choose how we want to evolve the quantum system from time $t$ to $t + \Delta t$.

\subsubsection{Algorithm}

Our goal is to deal with quantum systems having direct electron--electron interactions which warrants using a many-body quantum state to represent the electronic subsystem.
For this, we can draw from well-known many-body techniques such as the time-dependent Lanczos method~\cite{Park1986}, see \cref{fig:method_lanczos_mte}.
In the time-dependent Lanczos method, a Krylov subspace is constructed iteratively from repeated application of the Hamiltonian $\hat{H}_\text{MTE}(t)$ to the current quantum state $\ket{\psi(t)}$.
In the Krylov space, an effective Hamiltonian $\hat{H}_\mathcal{K}$ is found that can be used to evolve $\ket{\psi(t)}$ from time $t$ to $t + \Delta t$:
\begin{align}
    \ket{\psi(t + \Delta t)} = V_\mathcal{K}^\dagger \exp(-i \Delta t \hat{H}_\mathcal{K}) V_\mathcal{K} \ket{\psi(t)}\,,
\end{align}
with the transformation into the Lanczos basis $V_\mathcal{K}$.
$\hat{H}_\mathcal{K}$ is tridiagonal and has a much smaller dimension than the original Hamiltonian, allowing for easy diagonalization.

The accuracy of the Lanczos step is controlled both by increasing the dimension of the constructed Krylov subspace and by reducing  the Lanczos time-step $\Delta t_\mathrm{Lanczos}$~\cite{Manmana2005}.
As there exist readily available measurements for the error acquired in the time-dependent Lanczos step we can use an adaptive algorithm to fix the rate of error $\epsilon$ in the time-dependent Lanczos method, to an arbitrary value.
Choosing the Krylov dimension and $\Delta t_\mathrm{Lanczos}$ adaptively means that we can combine the algorithmic parameters for the time-dependent Lanczos method into the single parameter $\epsilon$.

Note that whenever the Lanczos step has to be decreased to ensure accuracy, resulting in $\Delta t_\mathrm{Lanczos} < \Delta t$, we still want to fix $\Delta t$ separately which is given by the alternating evolution of environment and subsystem in \cref{eq:splitting_method}.
In our concrete simulations, however, it is sufficient to allow for a maximum Krylov dimension $\approx 30$ to ensure that the time-evolution is accurate without a need to reduce $\Delta t_\mathrm{Lanczos}$; thus
$\Delta t_\mathrm{Lanczos}=\Delta t$ throughout this work.
Our implementation of time-dependent Lanczos is based on \cite{Haegeman2024}.

\subsection{Benchmark}

To demonstrate the performance of the Lanczos-MTE method we begin by comparing our results to those from Ref.~\cite{TenBrink2022}  for a non-interacting system, which is shown in~\cref{fig:mte_comparison_noninteracting}.
While in this case the less expensive single-particle formalism can be employed to evolve the quantum system, we show data that has been generated using the many-body Lanczos-MTE method.
For the example of $\mathcal{O}_\mathrm{CDW}$ we show that Lanczos-MTE and MTE produce equivalent results.
The inset in \cref{fig:mte_comparison_noninteracting} shows how the difference $\delta \mathcal{O}_\mathrm{CDW}$ between the different methods saturates around $\delta \mathcal{O}_\mathrm{CDW} \lesssim 10^{-2}$.
This is indeed expected as we consider an ensemble of $N_\mathrm{traj} = 4000$ trajectories:
Based on the statistical convergence analysis from~\cite{TenBrink2022} carried out for the same system size, the accuracy of observables in the ensemble, such as $\mathcal{O}_\mathrm{CDW}$, is $ \propto 1 / \sqrt{N_\mathrm{traj}}$.
The range of the imbalance $\mathcal{O}_\mathrm{CDW}(t)$ is $[-1, 1]$ and therefore, we can upper bound the statistical accuracy of $\mathcal{O}_\mathrm{CDW}(t)$ (and by extension also $\delta\mathcal{O}_\mathrm{CDW}(t)$) to be $2 / \sqrt{N_\mathrm{traj}} = 10 ^{-2}$.

\begin{figure}[t]
    \centering
    \includegraphics[width=246pt]{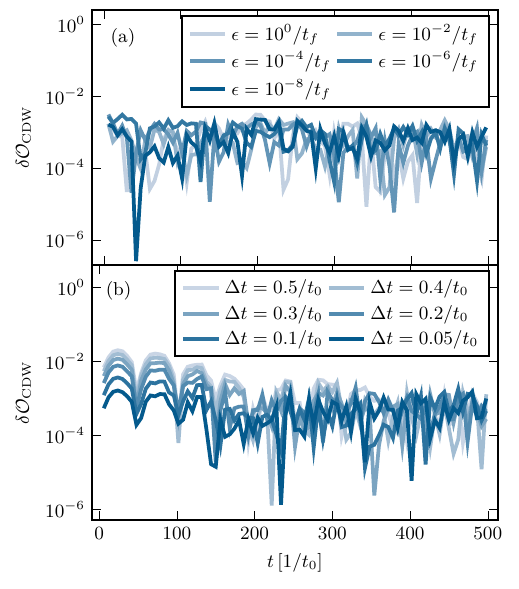}
    {\phantomsubcaption\label{fig:convergence_lanczos_errorrate}}
    {\phantomsubcaption\label{fig:convergence_lanczos_timestep}}
    
    \caption{
        Dependence of the Lanczos-MTE method on the
        control parameters time step $\Delta t$ and error rate $\epsilon$
        for an interacting system ($L=14$, $V=\qty{2}{\hopping}$, $\omega_0=\qty{0.1}{\hopping}$, $\gamma=\qty{0.4}{\hopping}$, $W=0$).
        We plot the deviation $\delta \mathcal{O}_\mathrm{CDW}$ when \subref{fig:convergence_lanczos_errorrate} varying $\epsilon$ at fixed $\Delta t = \qty{0.01}{\per\hopping}$, comparing to a reference set with $\epsilon = 10^{-10}/t_f$ and \subref{fig:convergence_lanczos_timestep} varying $\Delta t$ at fixed $\epsilon = 10^{-10} / t_f$, comparing to a reference set with $\Delta t = \qty{0.01}{\per\hopping}$.
        The results are averaged over $N_\mathrm{traj} = 4000$ trajectories and the final time $t_f = \qty{500} {\per\hopping}$.
    }
    \label{fig:lanczos_congvergence}
\end{figure}

In order to asses the accuracy of the Lanczos-MTE method we first analyze its dependence on $\epsilon$.
While the convergence behavior of the Lanczos step is generally well understood~\cite{Manmana2005}, we want to show how the accuracy of the Lanczos step influences the ensemble convergence in the MTE method.
In \cref{fig:convergence_lanczos_errorrate} we show the deviation $\delta\mathcal{O}_\mathrm{CDW} = | \mathcal{O}_{\mathrm{CDW}, \epsilon^*} - \mathcal{O}_{\mathrm{CDW},\epsilon}|$ where $\epsilon^*=10^{-10}/t_f$.
We observe that, again, the deviation $\delta \mathcal{O}_\mathrm{CDW}$ is never larger than $10^{-2}$.
We conclude that for all considered values of $\epsilon$ the deviation between the compared ensemble is dominated by the statistical error, indicating that Lanczos-MTE is well behaved with respect to the error rate $\epsilon$.

However, as we have already discussed in \cref{sec:mte}, the MTE approximation itself introduces a dependence on $\Delta t$.
To show that this dependence remains well-controlled within the additional approximations of Lanczos-MTE we plot the deviation $\delta \mathcal{O}_{\mathrm{CDW}} = | \mathcal{O}_{\mathrm{CDW},\Delta t^*} - \mathcal{O}_{\mathrm{CDW}, \Delta t}|$ for ${\Delta t ^* = \qty{0.01}{\per\hopping}}$ as a function of time and for different values of $\Delta t$ in \cref{fig:convergence_lanczos_timestep}.
We observe that for large $\Delta t$ and at early times, $\delta\mathcal{O}_{\mathrm{CDW}} < 10^{-2}$ does not hold. On the other hand, for long times, 
$\delta\mathcal{O}_{\mathrm{CDW}}< 10^{-2}$ appears to hold for all values of $\Delta t$ and for sufficiently small values of $\Delta t$, it holds everywhere. 
We observe that at long times $\delta \mathcal{O}_{\mathrm{CDW}}$ is essentially independent of $\Delta t$.
The fact that the accuracy of the ensemble average improves over time and does not deteriorate can be attributed to classically chaotic dynamics present in the single trajectories (see \cref{sec:appendix_chaos} for further details).

In summary, we have demonstrated that Lanczos-MTE reproduces electronic observables accurately by comparison to a standard MTE implementation for the non-interacting case using propagation of single-particle states. For the interacting case, the Lanczos errors can easily be made so small that
the dominating error is due the sampling of trajectories.

For local Hilbert spaces of dimension $2$, pure-state techniques such as Lanczos can reach system sizes as large as $L \simeq 30$ sites (see, e.g.,~\cite{Steinigeweg2014,Herbrych2012}). In combination with MTE, the sampling over trajectories
requires performing such calculations $N_{\mathrm{traj}}$ times, which adds computational costs that scale linearly in $N_{\mathrm{traj}}$.

\subsection{TEBD-MTE method}
\label{sec:DMRG-mte}

Similar to the Lanczos-MTE discussed in the previous section (\cref{sec:lanczos-mte}), we can use tensor network-based methods, such as the time-evolving block decimation (TEBD) method, for the quantum time evolution in \cref{eq:ehrenfest_eom3}. 
The evolution of the classical coordinates and momenta in \cref{eq:ehrenfest_eom3} is done in the same way as before. 
This combined TEBD-MTE method brings all the advantages of the TEBD method, such as simulating larger systems compared to the Lanczos-MTE and may allow applications to equilibrium and dynamical quantities. In \cref{fig:mte_comparison_noninteracting}, we have compared both the Lanczos-MTE and the TEBD-MTE method, which agree with each other within the error limits mentioned before.
Our TEBD implemtation utilizes the ITensor library \cite{itensor}.

\subsubsection{Algorithm}

In TEBD-MTE, the quantum many-body state $|\psi(t)\rangle$ of the electronic sub-system is described by a matrix product state (MPS).
The combined state of the whole system $(|\psi_{\{x_\ell, p_\ell\}}(t)\rangle, \{x_\ell,~ p_\ell\})$ is graphically represented in the bottom layer of \cref{fig:method_tebd_mte}. 
Figure~\ref{fig:method_mte_tensor} describes one graph element associated with a site $\ell$, which contains information about both the sub-systems, classical and quantum. 
Here, $A_{a_{\ell-1},a_\ell}^{\sigma_\ell}$ are the matrices of the MPS where $a_{\ell-1},a_\ell$ are the bond indices, and $\sigma_\ell$ is the physical index~\cite{Schollwoeck2011} and $x_\ell$ and $p_\ell$ represent the position and momentum coordinates of the classical phonon on site $\ell$. 

\begin{figure}[t]
    \centering
    
    \includegraphics[width=\linewidth]{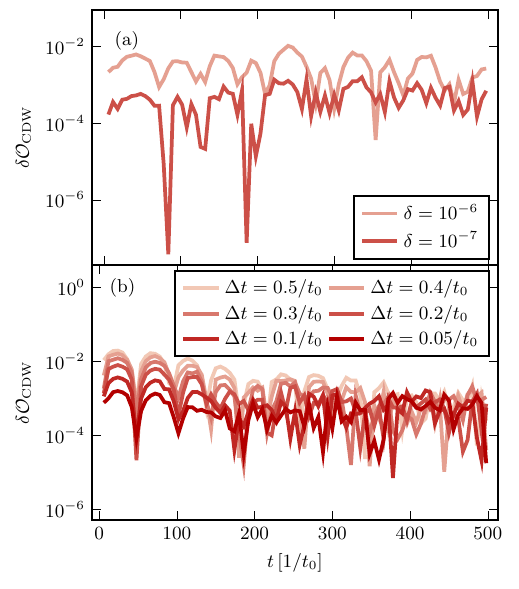}
    {\phantomsubcaption\label{fig:convergence_tebd_cutoff}}
    {\phantomsubcaption\label{fig:convergence_tebd_timestep}}
 \caption{
        Dependence of the TEBD-MTE method on the discarded weight $\delta$ and the time step $\Delta t$ in an interacting system ($L=14$, $V=\qty{2}{\hopping}$, $\omega_0=\qty{0.1}{\hopping}$, $\gamma=\qty{0.4}{\hopping}$, $W=0$).
        We plot the deviation $\delta \mathcal{O}_\mathrm{CDW}$ when \subref{fig:convergence_tebd_cutoff} varying $\delta$ at a fixed time step $\Delta t = \qty{0.01}{\per\hopping}$, comparing with a reference set using $\delta = 10^{-8}$ and \subref{fig:convergence_tebd_timestep} varying $\Delta t$ at fixed cutoff ${\delta = 10^{-8}}$, comparing with \changes{a} reference set using ${\Delta t = \qty{0.01}{\per\hopping}}$.
        The results are averaged over $N_\mathrm{traj} = 4000$ trajectories.
    }
    \label{fig:convergence_tebd}
\end{figure}

According to~\cref{eq:splitting_method}, $(|\psi_{\{x_\ell, p_\ell\}}(t)\rangle,\{x_\ell,p_\ell\})$ is time evolved by alternatively applying TEBD and classical updates as shown in \cref{fig:method_tebd_mte}. 
We use a second-order Trotter-Suzuki decomposition of $e^{-i\Delta t \hat{H}_\mathrm{MTE}}$ for the TEBD method as
\begin{align}\label{eq:ST}
    &\exp(-i\Delta t \hat{H}_\mathrm{MTE}) =
     \exp(-i\frac{\Delta t}{2} \hat{H}_\mathrm{MTE}^\mathrm{odd}) \\ 
     & \exp(-i\Delta t \hat{H}_\mathrm{MTE}^\mathrm{even})
     \exp(-i\frac{\Delta t}{2} \hat{H}_\mathrm{MTE}^\mathrm{odd})
     + \mathcal{O}(\Delta t^3)\nonumber \,.
\end{align}
Here, $\hat{H}_\mathrm{MTE} = \hat{H}_\mathrm{MTE}^\mathrm{odd} + \hat{H}_\mathrm{MTE}^\mathrm{even}$ and \begin{align}
    \hat{H}_\mathrm{MTE}^\mathrm{odd/even} = \sum_{j\in \mathrm{odd/even~bonds}}^{L-1} \hat{h}_j
\end{align}
are the parts of Hamiltonian $\hat{H}_\mathrm{MTE}$ which only operate on odd or even bonds of the lattice, respectively. Since the terms $\hat{h}_j$ in $\hat{H}_\mathrm{MTE}^\mathrm{odd/even}$ commute with each other, we can expand the terms in \cref{eq:ST} as \begin{align}
    \exp(-i\Delta t \hat{H}_\mathrm{MTE}^\mathrm{odd/even}) = \prod_{j\in \mathrm{odd/even~bonds}}^{L-1} \exp(-i\Delta t \hat{h}_j)\,.
\end{align}
Now, each term $\exp(-i\Delta t \hat{h}_j)$ in this equation can be applied to the MPS independently as a two-site gate (see \cref{fig:method_tebd_mte}). Following the application of a gate, a singular value decomposition (SVD) is performed to bring back the structure to MPS form. During the SVD process, we can truncate the smaller singular values $s_\alpha$ with a desired cutoff $\delta$. The number of singular values kept in this process defines the bond dimension $\chi$ of the MPS. The discarded weight is given by
\begin{align}
\delta = \sum_{\alpha=\chi+1}^{\alpha_{max}} s_\alpha^2\,.
\end{align}

As we know~\cite{Schollwoeck2011,Paeckel2019}, in the TEBD method, the error builds up with time, which depends on the time step in the Trotter-Suzuki decomposition and the truncation of the singular values during the singular value decomposition (SVD). 
The latter is related to the entanglement increase in a nonequilibrium problem and leads to an increase in the required resources. We can control the error of the TEBD step by decreasing the time step and the discarded weight.

\subsection{Benchmark}

\begin{figure}[t]
    \centering
    \includegraphics[width=\linewidth]{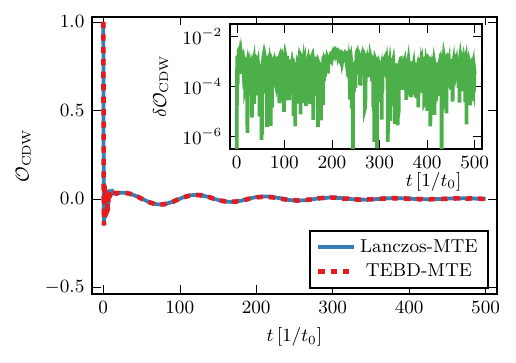}
    \caption{
        Comparison of the TEBD-MTE method (${\delta = 10^{-8}}$) and Lanczos-MTE method ($\epsilon = 10^{-8} / t_f$) in an interacting system ($L=14$, $V=\qty{2}{\hopping}$, $\omega_0=\qty{0.1}{\hopping}$, $\gamma=\qty{0.4}{\hopping}$) for the case without disorder ($W=0$).
        The inset shows the difference of the observable $\mathcal{O}_\mathrm{CDW}$ between the two methods as a function of time.
        Both methods employ a time step $\Delta t = \qty{0.01}{\per\hopping}$ and are sampled over $N_\mathrm{traj} = 4000$ trajectories. 
    }
    \label{fig:mte_comparison_interacting_nodisorder}
\end{figure}

\begin{figure}[t]
    \centering
    \includegraphics[width=\linewidth]{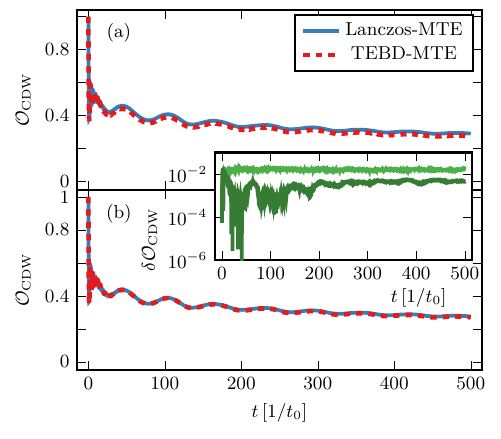}
    {\phantomsubcaption\label{fig:mte_comparison_interacting_disorder_lowsample}}
    {\phantomsubcaption\label{fig:mte_comparison_interacting_disorder_highsample}}
    \caption{
        Similar setup to \cref{fig:mte_comparison_interacting_nodisorder} for the disordered case ($W=\qty{8}{\hopping}$) for \subref{fig:mte_comparison_interacting_disorder_lowsample} $N_\mathrm{dis} = \num{200}$ and \subref{fig:mte_comparison_interacting_disorder_highsample} $N_\mathrm{dis} = \num{400}$ with $N_\mathrm{traj}=200$ for every disorder realization.
        The inset again shows the difference of the observable $\mathcal{O}_\mathrm{CDW}$ between the two methods as a function of time (green for $N_\mathrm{dis}=200$ and dark green for $N_\mathrm{dis}=400$).
        The parameters from \subref{fig:mte_comparison_interacting_disorder_lowsample} are used in the rest of this work while \subref{fig:mte_comparison_interacting_disorder_highsample} shows that the accuracy can be improved by increasing the number of disorder realizations.
    }
    \label{fig:mte_comparison_interacting_disorder}
\end{figure}

First, we analyze the performance of the TEBD-MTE method by decreasing the discarded weight $\delta$. The time evolution of the CDW state averaged over $4000$ trajectories for $L=14$, $V=\qty{2}{\hopping}$, $\omega_0=\qty{0.1}{\hopping}$, $\gamma = \qty{0.4}{\hopping}$, $W=0$ and $\Delta t = \qty{0.01}{\per\hopping}$ is shown in \cref{fig:convergence_tebd_cutoff}. 
Here, we plot the difference in $\mathcal{O}_{\rm{CDW}}(t)$ obtained from simulations with different values of the discarded weight, that is,
\begin{align}
\delta \mathcal{O}_\mathrm{CDW} = | \mathcal{O}_{\mathrm{CDW},\delta^*} - \mathcal{O}_{\mathrm{CDW},\delta}|\,,
\end{align}
where $\delta^* = 10^{-8}$ is the smallest $\delta$ considered.

As we can see from \cref{fig:convergence_tebd_cutoff}, $\delta \mathcal{O}_{\rm{CDW}}(t)$ decreases with smaller $\delta$ and $\delta \mathcal{O}_{\rm{CDW}}(t) \lesssim 10^{-2}$ for small values of $\delta$.
While evolving the system with different $\delta$, the $\chi_{\mathrm{max}}(t)$ at the central bond saturates to different values.
Here, $\chi_{\mathrm{max}}$ is the maximum of $\chi(t)$ in the ensemble of trajectories.
With decreasing $\delta$, $\chi_{\mathrm{max}}$ saturates at a higher number such as $\chi_{\mathrm{max}} = 84,~119,$ and $128$ for $\delta = 10^{-6},~ 10^{-7},$ and $ 10^{-8}$, respectively.
Note that the maximum $\chi$ at the center of the system of $L=14$ is $2^{L/2} = 128$ if we were to carry out the numerical calculation without truncation.

Similarly, we study the dependence on $\Delta t$
in \cref{fig:convergence_tebd_timestep}. 
Note that here we use the same time step $\Delta t$ in both \cref{eq:ST} for the Trotter-Suzuki decomposition and \cref{eq:splitting_method} for the splitting method.
We consider the same Hamiltonian parameters with the smallest $\delta$ considered here ($\delta = 10^{-8}$) and $N_{\rm{traj}} = 4000$ and plot $\delta \mathcal{O}_\mathrm{CDW} = | \mathcal{O}_\mathrm{CDW}(\Delta t^*) - \mathcal{O}_\mathrm{CDW}(\Delta t)|$ for different $\Delta t$ with respect to the smallest $\Delta t = \Delta t^* = \qty{0.01}{\hopping}$. The figures show that again, $\delta \mathcal{O}_{\rm{CDW}} \lesssim 10^{-2}$ and hence, the statistical errors dominate.

Using the smallest $\Delta t$ and $\delta$ considered above for the TEBD-MTE method, we compare the time evolution of the system with the Lanczos-MTE method with the smallest $\Delta t$ and $\epsilon$ discussed in the previous section \cref{sec:lanczos-mte}, now for an \emph{interacting} system. 
We consider the model parameters $L=14$, $V=\qty{2}{\hopping}$, $\omega_0=\qty{0.1}{\hopping}$, $\gamma = \qty{0.4}{\hopping}$. 
The time evolution of a CDW state in terms of $\mathcal{O}_{\rm{CDW}}(t)$ is shown in \cref{fig:mte_comparison_interacting_nodisorder,fig:mte_comparison_interacting_disorder} for $W=0$ and $W=\qty{8}{\hopping}$, respectively. 
The evolution of $\mathcal{O}_{\rm{CDW}}(t)$ agrees well between the methods for both cases. The difference $\delta \mathcal{O}_{\rm{CDW}}(t)$ between the methods, shown in the inset of both panels, stays below $O(10^{-2})$ for the case $W=0$.
For the case $W=\qty{8}{\hopping}$ we show in \cref{fig:mte_comparison_interacting_disorder_lowsample,fig:mte_comparison_interacting_disorder_highsample} how in the cases with disorder sampling the MTE accuracy can be improved by increasing the amount of disorder samples.
Since we aim at qualitative results in this work, we choose $N_{\text{dis}}=200$ for the rest of this work for $W>0$ and more than one electron.

In the simulations presented so far, we use TEBD-MTE on small systems where a direct comparison to MTE-Lanczos is possible yet the bond dimension truncation is still irrelevant.
For the main application presented in this work, the decay of CDW order in disordered interacting systems, MTE-Lanczos is the better suited method since reaching long times is crucial and therefore, Lanczos-MTE will be solely used in \cref{sec:mbl}.
TEBD, as all time-dependent MPS methods do, is ultimately limited to short times in situations with a significant entanglement growth.

We envision several interesting applications of TEBD-MTE beyond the cases covered here. Generally, one of the most important applications of time-dependent MPS methods is the calculation of equilibrium spectral functions and other dynamical properties. These are usually cases with a mild entanglement increase, even at finite temperature.
While the regime of intermediate to large values of $\omega_0$ have been covered with full-quantum MPS methods~\cite{Jansen2020,Jansen2023},
the adiabatic regime of optical phonons with interacting electrons is mostly unexplored with MPS methods, which also applies to \changes{acoustic} phonons. In the long run, using hybrid MPS-trajectory-based methods in DMFT might be an interesting direction as well. In order to illustrate that TEBD-MTE is well behaved concerning its convergence properties on larger systems, we present further results in Appendix~\ref{sec:L28}.

To summarize, the two methods, Lanczos-MTE and TEBD-MTE, produce consistent results with very good quantitative agreement.
For the chosen model and simulation parameters, the accuracy is limited
by the statistical errors due to sampling Ehrenfest trajectories.

\section{Stability of an Anderson insulator}
\label{sec:anderson}

In this section, we discuss the case of non-interacting electrons ($V=0$) with finite disorder. 
When the electronic sub-system is isolated from the phonons, the system exhibits Anderson localization for any finite disorder strength in one dimension~\cite{Kramer1993}. In this section we study the stability of the Anderson localization in the presence of classical optical phonons. 

\changes{Our expectation is that} for small $\gamma$ and $\omega_0$, MTE will accurately reproduce the electronic observables even of the original quantum model. 
However, we will go beyond the regime of $\gamma \ll \unit{\hopping}$ addressing the question of the stability of Anderson laocalization against a \emph{classical} bath for arbitrary values of $\gamma$, regardless of whether the quantum version is accurately described or not. 
The same philosophy will apply in \cref{sec:mbl}. This section serves to set the stage for the analysis of systems with electronic interactions; so long as $V=0$, no many-body treatment of the electronic sector is necessary and instead, we use single-particle orbitals in the MTE implementation as in~\cite{TenBrink2022}. 
We first study the spreading of a single electron and then move on to the decay of the CDW initial state. 

\subsection{Spreading of a single electron}

\begin{figure}[t]
    \centering 
    \includegraphics[width=\linewidth]{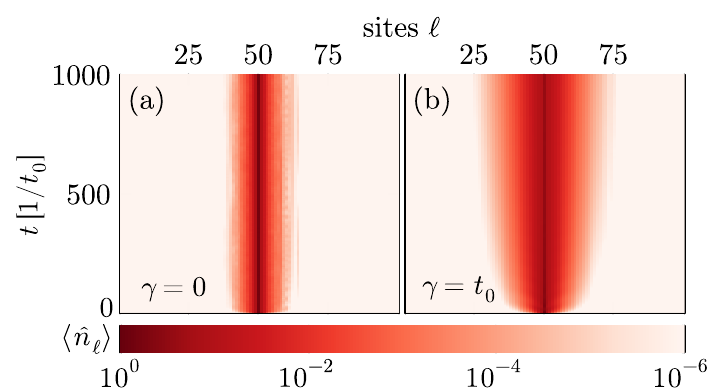}
    {\phantomsubcaption\label{fig:density_evolution_gamma=0}}
    {\phantomsubcaption\label{fig:density_evolution_gamma=1}}

    \caption{
        Time-evolved density profile $\langle \hat n_\ell(t)\rangle $ of a particle initially ($t=0$) localized on a single lattice site $\ell_\changes{0}=50$ on a lattice with $L=100$ sites and with disorder (${W=\qty{8}{\hopping}}$).
        We show the \subref{fig:density_evolution_gamma=0} uncoupled case $\gamma=0$, exhibiting Anderson localization, and \subref{fig:density_evolution_gamma=1} a system with a large electron--phonon coupling $\gamma=\unit{\hopping}$.
        Further parameters are $\omega_0 = \qty{0.1}{\hopping}$ and the data was generated using the Lanczos-MTE routine with $\Delta t = \qty{0.01}{\per\hopping}$, $\epsilon = 10^{-9} / t_f$, and $N_\mathrm{dis} = 5000$ realizations for $N_\mathrm{traj}=200$ trajectories each.
    }
    \label{fig:density_evolution}

\end{figure}

\begin{figure}[t]
    \centering    
    {\centering\includegraphics[width=\linewidth]{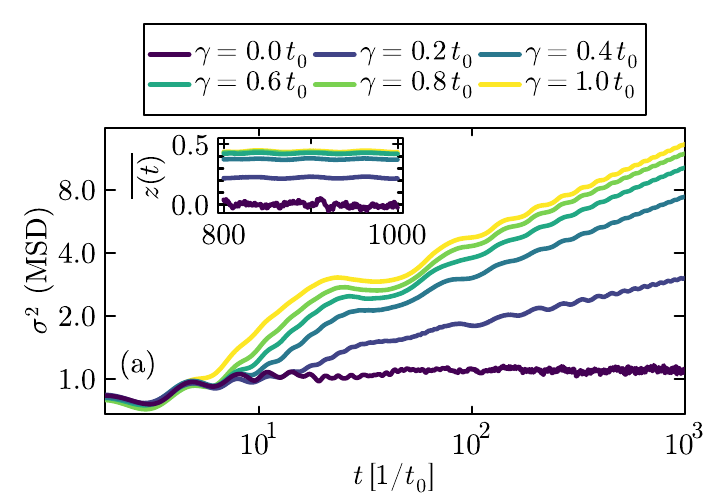}\phantomsubcaption\label{fig:oneparticle_msd}}
    {\centering\includegraphics[width=\linewidth]{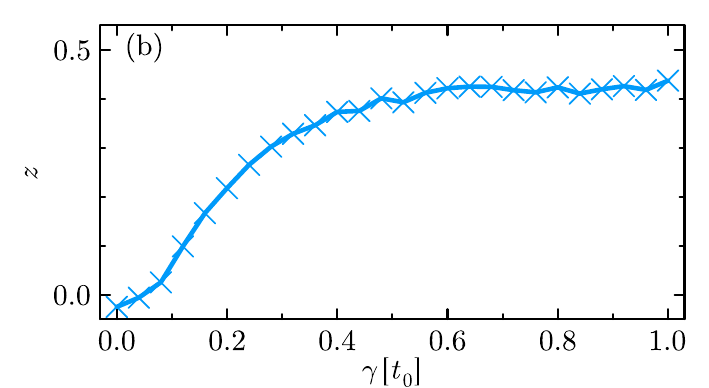}\phantomsubcaption\label{fig:oneparticle_exponent}}

    \centering
    \caption{
        Spreading of a single particle in the disordered system coupled to classical phonons. 
        For a few values of $\gamma$ we show in \subref{fig:oneparticle_msd} the MSD for the particle as a function of time and in the inset the time-dependent dynamical exponent $\bar z(t)$, averaged over a time window of $t^* = \SI{800}{\per\hopping}$ (see \cref{eq:time_averaged_dynamicalexp}). 
        In \subref{fig:oneparticle_exponent} the extracted estimates for the dynamical exponent $z$ is shown. 
        The model parameters are $L=100$, $\omega_0 = \qty{0.1}{\hopping}$, and $W=\qty{8}{\hopping}$ and the particle is initially localized at lattice site $\ell_\changes{0} = 50$.
        The data are generated using an MTE routine using exact diagonalization in the single-particle picture to propagate the quantum-state with $N_\mathrm{dis}=2000$ realizations for $N_\mathrm{traj} = 200$ trajectories each and $\Delta t = \qty{0.01}{\per\hopping}$.
    }
    \label{fig:oneparticle_gammasweep}
\end{figure}

\begin{figure}[t]
    \centering    
    {\centering\includegraphics[width=\linewidth]{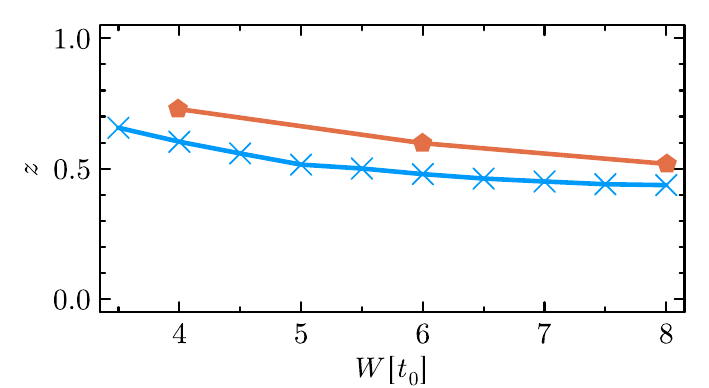}\phantomsubcaption\label{fig:oneparticle_exponent_disorder}}

    \centering
    \caption{
        Similar setup to \cref{fig:oneparticle_gammasweep} with $\gamma = \unit{\hopping}$ held fixed while $W$ is varied.
        We show the dependency of the dynamical exponent $z$ on disorder strength obtained from MTE. 
        Our estimate of the dynamical exponent $z$ is averaged over a large time window $t^* = \SI{800}{\per\hopping}$.
        The solid symbols are data from a full quantum-treatment of electrons coupled to dispersive hard-core bosons for $\omega_0=\gamma=\unit{\hopping}$ from~\cite{Prelovsek2018}, with a hard-core boson hopping amplitude $t_b=\qty{0.4}{\hopping}$.
    }
    \label{fig:oneparticle_disordersweep}
\end{figure}

 Anderson localization implies no transport at all in one dimension~\cite{Evers2008}, so any persistent spreading of a particle can be interpreted as a breakdown of Anderson localization.
In \cref{fig:density_evolution} we show how a particle on a lattice of $L=100$ sites and disorder strength $W = \qty{8}{\hopping}$ spreads out in the system over time. 
The particle is initially localized at site $\ell_\changes{0}=50$.
For $\gamma = 0$ we clearly see that the particle stays localized  as expected. 
In contrast, when $\gamma=\unit{\hopping}$, the particle continuously spreads out with a characteristic cone whose shape is related to the specific transport properties of the system.

In order to classify the dynamics, we extract the dynamical transport exponent $z$ defined from the asymptotic behavior of the \emph{mean-square-displacement} (MSD, $\sigma^2$) with $\sigma^2(t) \propto t^z$ for large $t$.
The MSD can be readily calculated from:
\begin{align}
    \label{eq:mean_square_displacement}
    \sigma^2(t) = \sum_{\ell=1}^L (\ell - \ell_0)^2 \langle \hat{n}_\ell(t)\rangle\,,
\end{align}
where $\ell_0$ is the site where the electron is initially localized on the lattice.

To extract the dynamical exponent $z$ from the MSD we first find the time-dependent dynamical exponent $z(t)$ using the method of the logarithmic derivative where 
\begin{align}
    \label{eq:logarithmic_derivative}
    z(t) = \dv{\log(\sigma^2(t))}{\log(t)} \approx t\frac{\delta \log(\sigma^2(t))}{\delta t} \,.
\end{align}
 We use a simple finite difference numerical differentiation approach using $\delta t = \qty{2}{\per\hopping}$.
Lastly, we average $z(t)$ over a large time window of size $t^* = \qty{800}{\per\hopping}$ to remove coherent fluctuations present in the data due to oscillations in the classical phonon sector.
The time-averaged dynamical exponent is therefore
\begin{align}
    \label{eq:time_averaged_dynamicalexp}
    \overline{z(t)} = \mean\limits_{t - t^* < t^\prime \le t} z(t^\prime)\,.
\end{align}
To extract an estimate for the asymptotic dynamical exponent $z$ we choose $z = \overline{z(t_f)}$ with $t_f$ being the final time.

We show the MSD and the extracted dynamical exponents as a function of the electron--phonon coupling $\gamma$ in \cref{fig:oneparticle_msd,fig:oneparticle_exponent}, respectively.
In \cref{fig:oneparticle_exponent} we observe  a monotonic increase of $z$ as a function of $\gamma$.
At around $\gamma \gtrsim \qty{0.1}{\hopping}$, the dynamical exponent increases sharply until nearly saturating for $\gamma \sim \qty{0.5}{\hopping}$ slightly below $z \sim 0.5$.
As $z=1$ would indicate diffusive transport, the results indicate that there is a broad subdiffusive regime induced by the coupling to the classical phonon bath.

Since the extraction of $z$ is less reliable at small values of $W/\unit{\hopping}$ due to larger finite-size effects, we focus on the regime of $W\gtrsim \qty{4}{\hopping}$.
Observing that $\sigma \ll L$ in \cref{fig:oneparticle_msd} we infer that the particle has not yet spread over the whole system and therefore finite-size effects are expected to be insignificant.
In \cref{fig:oneparticle_disordersweep}, the dynamical exponent increases monoton\vanishchanges{e}ously as a function of $W$ in the parameter regime studied here.
Since $z <1$, we observe subdiffusive dynamics. 
As $\gamma\to 0$, we recover localization. 
If $W$ were to be sent to zero at finite $\gamma$, we expect diffusion since the model is nonintegrable and \changes{fulfills the} eigenstate thermalization hypothesis~\cite{Jansen2020}, even though this is not firmly established for the Holstein model with one electron~\cite{Schoenle2021} (see also~\cite{Bai2018,Carof2019} for an independent trajectory based study using fewest-switches surface-hopping).
Recent work on the Holstein chain using MTE and quantum typicality \cite{Miladic2025,Mitric2025,Mitric2025a}  seems consistent with diffusion. Generally,
for a single charge carrier coupled to an Anderson insulator \cite{Sierant2023}, slow dynamics is often observed.

Interestingly, the qualitative and quantitative dependence of $z$ on $W$ is very similar to that obtained for $\omega_0 = \gamma = \unit{\hopping}$ obtained in~\cite{Prelovsek2018} from a numerically exact approach studying an electronic system coupled to dispersive hard-core bosons (see the circles in \cref{fig:oneparticle_disordersweep}).

In summary, we observe delocalization and subdiffusive electronic transport due to the coupling to classical phonons. Going to more electrons and interacting system is expected to accelerate that delocalization which we will capture by \changes{studying} the decay of $\mathcal{O}_{\text{CDW}}$.

\subsection{Many electrons, \texorpdfstring{$V=0$}{V=0}: Decay of CDW order}
\begin{figure}[t]
    \centering
    {\centering\includegraphics[width=\linewidth]{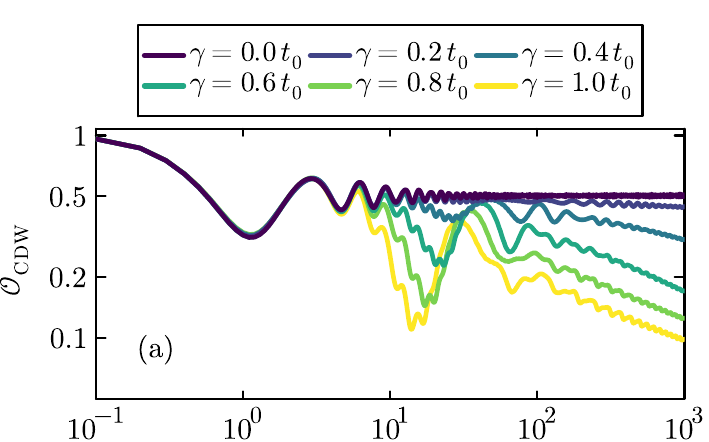}\phantomsubcaption\label{fig:noninteracting_gammasweep_imbalance}}
    {\centering\includegraphics[width=\linewidth]{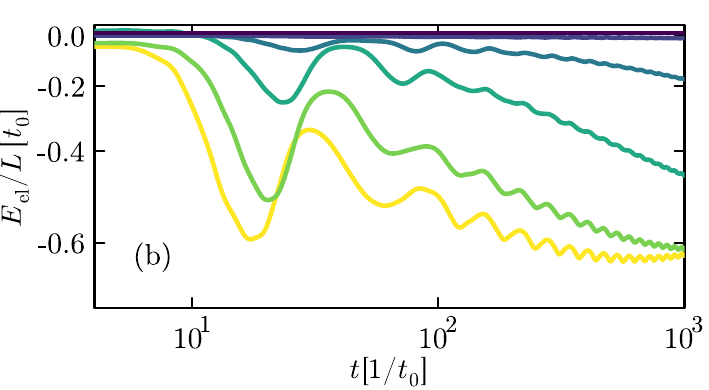}\phantomsubcaption\label{fig:noninteracting_gammasweep_energy}}

    \centering
    \caption{
        Decay of the CDW state in a non-interacting, disordered system as captured by the \subref{fig:noninteracting_gammasweep_imbalance} order parameter $\mathcal{O}_\mathrm{CDW}$ and \subref{fig:noninteracting_gammasweep_energy} electronic energy $E_\mathrm{el}$ for different values of $\gamma$ ($L=100$, $V=0$, $\omega_0 = \qty{0.1}{\hopping}$, $W=\qty{8}{\hopping}$).
        The data was generated using the Lanczos-MTE algorithm with $\Delta t = \qty{0.01}{\per\hopping}$, $\epsilon = 10^{-10} / t_f$ and $N_\mathrm{dis}=200$ with $N_\mathrm{traj} = 200$.
    }
    \label{fig:noninteracting_gammasweep}
\end{figure}

We consider a system of $L=100$ and a large disorder strength $W=\qty{8}{\hopping}$ at half filling, i.e., $N=L/2$ electrons. 
Starting with a CDW state, we time evolve the initial state for different values of the electron--phonon coupling~$\gamma$. 

The evolution of \vanishchanges{the} $\mathcal{O_{\rm{CDW}}}$ averaged over $4000$ trajectories is shown in \cref{fig:noninteracting_gammasweep_imbalance}. For $\gamma =0$, $\mathcal{O_{\rm{CDW}}}(t)$ saturates quickly and stays finite at all times, implying Anderson localization. 
By making the coupling to the phonons finite, though, localization is not stable anymore. The imbalance $\mathcal{O_\mathrm{CDW}}$ decays following a power law. We will discuss the exponents of the power-law decay together with the interacting case in \cref{sec:mbl}.

We also calculate the energy of the electronic subsystem 
\begin{align}
E_{\mathrm{el}}(t) = \mel{\psi(t)}{(\hat{H}_\mathrm{kin} + \hat{H}_\mathrm{in} + \hat{H}_\mathrm{dis})}{\psi(t)}\,,
\end{align} 
shown in \cref{fig:noninteracting_gammasweep_energy}. 
When the system is localized (${\gamma=0}$), $E_{\rm{el}}$ is time-independent, but for finite values of $\gamma$, $E_{\rm{el}}$ decreases implying an energy exchange between the electronic subsystem and the phonons, which leads to the destabilization of the Anderson localized system.

\section{Stability of MBL}
\label{sec:mbl}

\begin{figure}
    \centering
    \includegraphics[width=\linewidth]{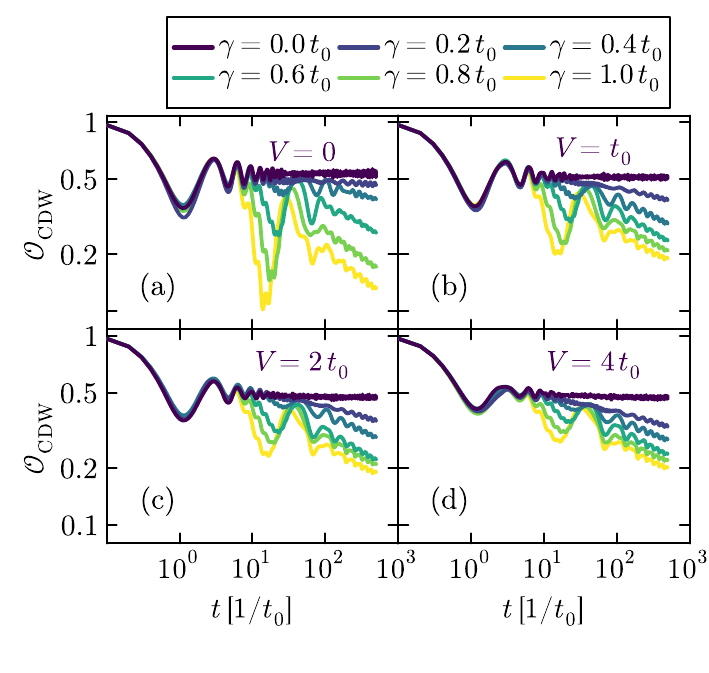}%
    {\phantomsubcaption\label{fig:interacting_gammasweep_V=0}}
    {\phantomsubcaption\label{fig:interacting_gammasweep_V=1}}
    {\phantomsubcaption\label{fig:interacting_gammasweep_V=2}}
    {\phantomsubcaption\label{fig:interacting_gammasweep_V=4}}
    \vspace{-2.2em}

    \caption{
        Decay of the CDW order parameter in an interacting, disordered system for different values of $\gamma$ ($L=14$, $\omega_0 = \qty{0.1}{\hopping}$, $W=\qty{8}{\hopping}$).
        Electron-electron interactions are investigated in the range of $V/\unit{\hopping}=0,1,2,4$ in panels (\subref{fig:interacting_gammasweep_V=0}-\subref{fig:interacting_gammasweep_V=4}), respectively.
        The data are generated using the Lanczos-MTE algorithm with $\Delta t = \qty{0.01}{\per\hopping}$, $\epsilon = 10^{-10} / t_f$, and $N_\mathrm{dis}=200$ with $N_\mathrm{traj} = 200$.
    }
    \label{fig:interacting_gammasweep}
\end{figure}

When we introduce interactions between electrons, when uncoupled to phonons, finite systems exhibit many-body localization (MBL) for large disorder strengths~\cite{dAlessio2016,Gogolin2016,Deutsch2018,Mori2018}. 
The stability of MBL in the presence of a bath has been studied in various scenarios both theoretically~\cite{Nandkishore2015,Sarang2017,Luitz2017,Katharine2017,Pollmann2020,Abanin12021} and experimentally~\cite{Bloch2019,Greiner2023}. In some cases, the bath delocalizes MBL~\cite{Sarang2017,Luitz2017,Pollmann2020,Abanin12021} (on accessible time scales and system sizes), whereas in other cases, MBL can even localize the bath~\cite{Nandkishore2015,Katharine2017} via a proximity effect. In this section we study the stability of finite-size MBL when coupled to a classical bath consisting of classical dispersion-less phonons.

\subsection{Decay of charge-density-wave states}

We consider a large disorder strength $W=\qty{8}{\hopping}$ and similar to the previous case of $V=0$ from \cref{sec:anderson}, we time evolve the CDW state (see \cref{eq:cdw_state}) for different values of $\gamma$~\cite{Abanin2023}. 
Here, the Lanczos-MTE is used for $L=14$ and the evolution of $\mathcal{O}_\mathrm{CDW}(t)$ is averaged over $4000$ trajectories. 

The results are shown in \cref{fig:interacting_gammasweep} for $L=14$ and $V/\unit{\hopping}=0,1,2,4$. 
We see that the system is localized when $\gamma = 0$ as expected for MBL on finite system sizes. 
For finite $\gamma$, $\mathcal{O}_\mathrm{CDW}(t)$ decays following power-laws similar to the instability of the Anderson-localized case (compare \cref{fig:noninteracting_gammasweep_imbalance}) which is overlayed by a persistent oscillation up to long times. 
The observation of long-time oscillatory behavior is typical in MTE simulations due to overcoherent effects as discussed in \cref{sec:strengths_and_weaknesses}.

\subsection{Power-law decay}

As a significant feature, we observe that $\mathcal{O}_\mathrm{CDW}$ decays asymptotically \changes{with a power law for all values of $\gamma$ considered}.
We observe, that adding interactions to the electron system leads to a faster decay of $\mathcal{O}_{\text{CDW}}$ at small $\gamma$ as compared to the $V=0$ case.
At large $\gamma$ this effect is strongly reduced:
In fact, we observe that the decay of $\mathcal{O}_\mathrm{CDW}$ is much reduced when $V>0$ for large $\gamma$.
We will provide a qualitative explanation of these trends in the next section.

In the single-particle case, we clearly observe subdiffusion at finite $\gamma$ and $W>0$, with dynamical exponents $z\leq 0.5$ for large $W=\qty{8}{\hopping}$. 
While it is not obvious that this behavior carries over to the many-body case, there are nevertheless suggestions for a connection between the dynamical exponent $z$ and the change in the behavior of $\mathcal{O}_{\text{CDW}}$ (or more generally, density autocorrelations)~\cite{Luitz2017a}. 
This would suggest sub-diffusive dynamics even in the many-body case.
Whether such behavior can persist up to asymptotically long times is an open question left for future research.

\subsection{Discussion and Interpretation}
\label{sec:discussion}

In the previous sections we have discussed how the coupling to classical phonons leads to delocalization in disordered systems, both for $V=0$
and $V>0$.
There is a qualitative difference in the regime of $\gamma\ll\unit{\hopping}$, where interacting systems relax faster and $\gamma \sim \unit{\hopping}$ where larger interactions inhibit relaxation.

In this section we will provide a discussion of these qualitative observations. 
Generally, without the MTE approximation, a finite $\gamma$ introduces interactions between the electrons.
One may therefore attempt to integrate out the phonons, mapping everything to an effective electronic model.
This, however, will be insufficient to lead to the observed fast decay of $\mathcal{O}_{\text{CDW}}$ as an actual energy transfer to the phononic subsystem appears to be crucial for delocalization.
The rate of that transfer is controlled by $1/\gamma$. 
Nonetheless, at weak $\gamma$, it is not surprising that larger electronic interactions accelerate the decay of $\mathcal{O}_\text{CDW}$ as this trend is already seen in the absence of a coupling to any phonons~\cite{Luitz2016a,Luitz2017a}. 

In the regime of large $\gamma >\unit{\hopping}$, without resorting to the classical and mean-field approximations of MTE, increasing $\gamma$ will make the physics more local and in the extreme case produce small polarons.
These quasiparticles have a smaller bandwidth than the bare electron and will therefore generally see a large effective disorder potential (for a more detailed discussion, see~\cite{Bronold2002,Bronold2004}).
This suggests that at fixed disorder strength $W$, there are two competing trends: increasing $\gamma$ favors the energy transfer to the phonons and hence relaxation of density inhomogeneities but at the same time leads to less mobile particles and thus in itself to localization.
Our numerical observations from \cref{fig:noninteracting_gammasweep,fig:interacting_gammasweep} are consistent with this picture.

There is another path towards understanding the delocalization due to the Holstein coupling to phonons that is particularly evident in the MTE approximation. 
The fluctuating phonon degrees of freedom couple directly to the electronic density $\langle \hat n_\ell \rangle$ and hence themselves induce disorder potential\changes{s} dynamically~\cite{DiSante2017}.
On average, over time and trajectories, these average out, yet can cancel out the static disorder potential instantaneously.

\begin{figure}
    \centering
    {\centering\includegraphics[clip, trim={{3.3\linewidth} {0\linewidth} {0.0\linewidth} {1.\linewidth}}, width=1.\linewidth]{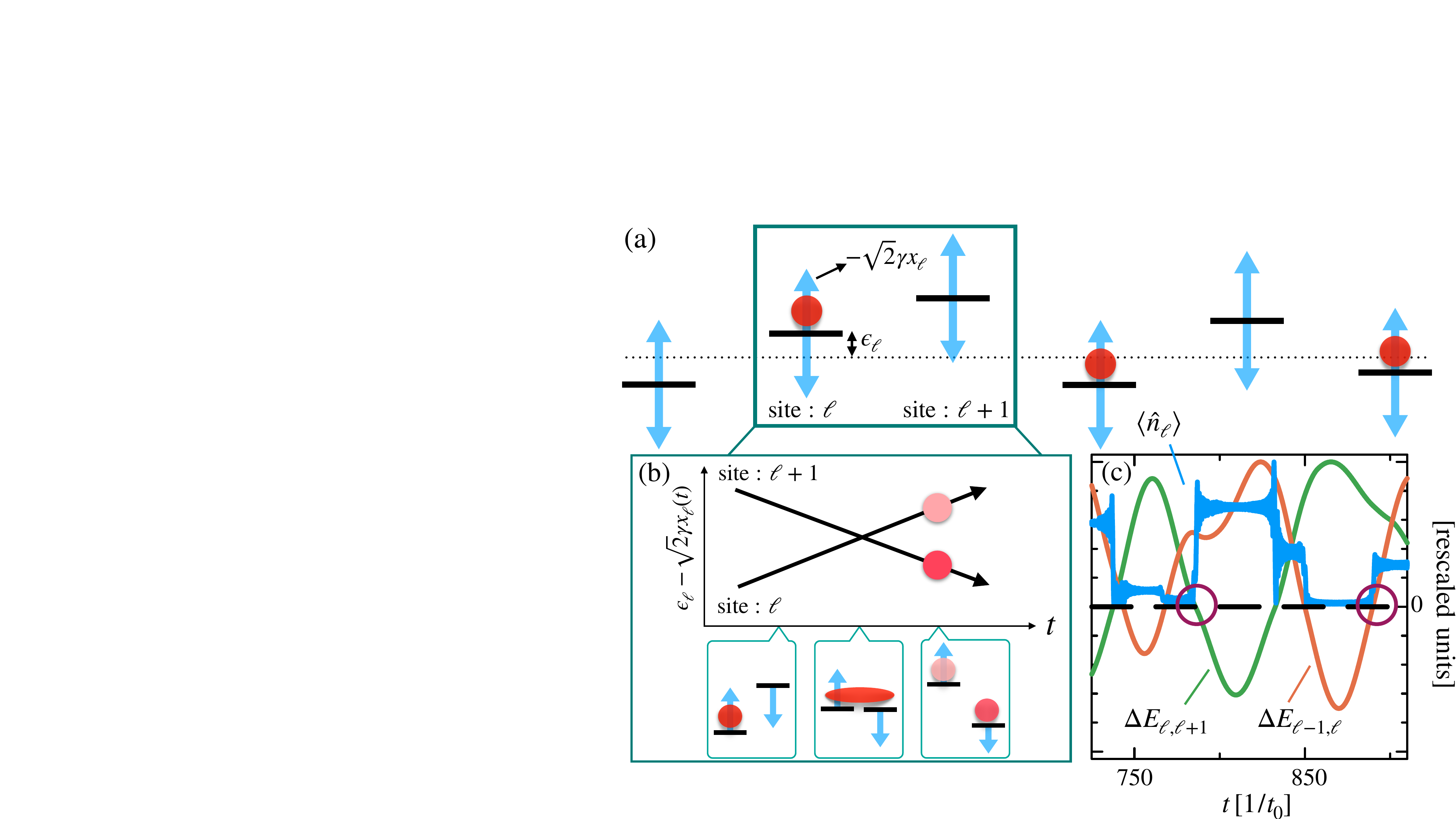}}
    {\phantomsubcaption\label{fig:process_chain}}
    {\phantomsubcaption\label{fig:process_transition}}
    {\phantomsubcaption\label{fig:process_data}}
    \caption{
        Sketches explaining relaxation processes induced by coupling to classical phonons in a single trajectory/realization.
        \subref{fig:process_chain} Examplary setup of the chain for the CDW initial state. 
        On each site $\ell$ there is a random disorder field $\epsilon_\ell$. 
        On top of this static disorder the electron--phonon coupling imposes a dynamical quasi-disorder field, depending on the classical coordinates $x_\ell$.
        \subref{fig:process_transition} Local relaxation process of an electron induced by the dynamical disorder fluctuations.
        As the energy difference between neighboring sites becomes small the electron delocalizes over both sites on a much shorter timescale than the fluctuations in the disorder, therefore resonant tunneling to \vanishchanges{to }the next site is possible.
        \subref{fig:process_data} Sketch created from real data of a single realization for the parameters of \cref{fig:oneparticle_gammasweep}.
        We choose a large value of $\gamma = \qty{5}{\hopping}$ to exaggerate the process that we want to highlight. 
        On a selected lattice site $\ell=46$ and over a selected time frame, we show the site occupation $\langle \hat{n}_\ell \rangle$ and the on-site energy differences $\Delta E_{\ell-1, \ell}$ ($\Delta E_{\ell, \ell+1}$) induced by the time-dependent disorder between the left (right) lattice sites.
        As a guide to the eye, we have added circles at some of the points in time where $\langle \hat{n}_\ell \rangle$ jumps due to $\Delta E_{\ell-1, \ell} = 0$ or $\Delta E_{\ell, \ell+1} = 0$.
        The values of $\langle \hat{n}_\ell \rangle$ and $\Delta E_{\ell-1, \ell}$ ($\Delta E_{\ell, \ell+1}$) are rescaled for better visibility by setting the maximum value of each curve over the inspected interval to one.
    }
    \label{fig:process}
\end{figure}

\begin{figure}[ht]
    \centering
    \includegraphics[width=\linewidth]{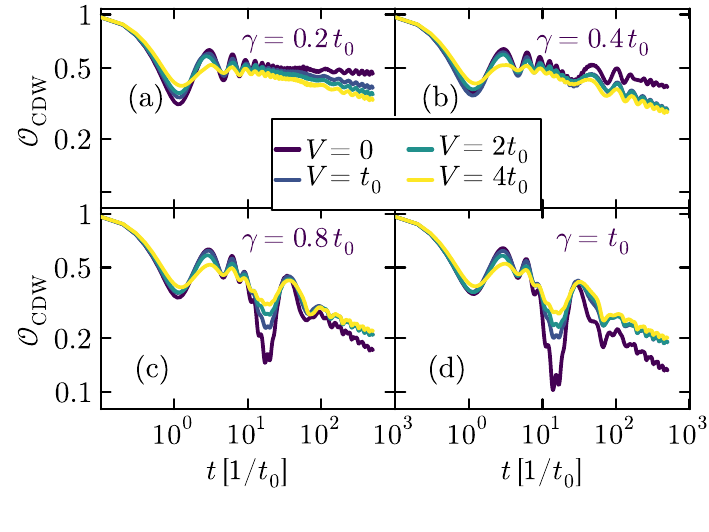}
    {\phantomsubcaption\label{fig:raw_cdw_decay_interacting_fixed_gamma=0.2}}
    {\phantomsubcaption\label{fig:raw_cdw_decay_interacting_fixed_gamma=0.4}}
    {\phantomsubcaption\label{fig:raw_cdw_decay_interacting_fixed_gamma=0.8}}
    {\phantomsubcaption\label{fig:raw_cdw_decay_interacting_fixed_gamma=1}}
    \caption{
        Decay of the CDW order parameter in an interacting, disordered system for different values of $V$ ($L=14$, $\omega_0 = \qty{0.1}{\hopping}$, $W=\qty{8}{\hopping}$).
        Electron-phonon couplings are investigated in the range of $\gamma/\unit{\hopping}=0.2,0.4,0.8,1$ in panels \subref{fig:raw_cdw_decay_interacting_fixed_gamma=0.2}-\subref{fig:raw_cdw_decay_interacting_fixed_gamma=1}, respectively.
        The data are generated using the Lanczos-MTE algorithm with $\Delta t = \qty{0.01}{\per\hopping}$, $\epsilon = 10^{-10} / t_f$, and $N_\mathrm{dis}=200$ with $N_\mathrm{traj} = 200$.
    }
    \label{fig:raw_cdw_decay_interacting_fixed_gamma}
\end{figure}

This leads to a classical picture of resonant local relaxation processes. 
We consider a single trajectory in the limit of $\gamma > \unit{\hopping}$. 
The total effective disorder at lattice site $\ell$ at time $t$ is $\epsilon_{\mathrm{total},\ell} = \epsilon_\ell - \sqrt{2}\gamma x_\ell(t)$.
We observe that whenever $\epsilon_{\mathrm{total},\ell}$ on two neighboring sites is nearly equal, the electron can resonantly tunnel into a previously empty site.
A sketch is shown in \cref{fig:process_transition}.
Since the \changes{electronic} sector undergoes much faster dynamics than the disorder landscape at small $\omega_0$, changes in the potential landscape due to the classical phonons may enable local relaxation processes.
The presence of repulsive interactions detunes these direct transition, hence at large $\gamma$, increasing $V$ leads to slower relaxation.
This is illustrated in \cref{fig:raw_cdw_decay_interacting_fixed_gamma}.

Lastly, we show an illustration based on actual numerical data in the large coupling limit $\gamma = \qty{5}{\hopping}$ in \cref{fig:process_data}, to support our previous arguments.
For a single realization and a selected lattice site and time frame, we show the on-site energy difference to the left (right) lattice site $\Delta E_{\ell, \ell+1} = E_\ell - E_{\ell+1}$ ($\Delta E_{\ell-1, \ell} = E_{\ell-1} - E_{\ell}$) and the particle occupation of the selected site $\langle \hat{n}_\ell(t)\rangle$ (rescaled for easier visualization since we are representing $\langle\hat n_\ell \rangle$ and $\Delta E$ in the same panel). 
We observe that $\langle \hat{n}_\ell(t) \rangle$ nearly resembles a step function in time where changes in the value coincide with $\Delta E_{\ell, \ell+1} = 0$ or $\Delta E_{\ell-1, \ell} = 0$ indicating the presence of local resonant relaxation processes\footnote{
We remark that the Wigner function represented by the ensemble of trajectories cannot be interpreted as a probability distribution and therefore there is no way to assign physical significance to a single trajectory in the ensemble. 
However, by considering a regime where MTE does not reproduce quantum mechanical results we also have effectively given up on insisting that the ensemble represents a Wigner function corresponding to a quantum mechanical state.
Therefore, there is no problem in the analysis of single realizations.}.

In the regime where $\gamma \sim \unit{\hopping}$ the processes outlined above are harder to identify.
Nevertheless, we hypothesize that the qualitative picture obtained from $\gamma\gg \unit{\hopping}$ remains valid for the entire regime of values of $\gamma$. As $\gamma$ decreases, the dynamical contributions to the local potential will increasingly often not be able to overcome the static disorder, eventually blocking this local resonant relaxation. 
A remaining potential relaxation channel is longer-range tunneling akin to variable-range hopping, which we do not further analyze in our data.

To summarize our discussion of the CDW relaxation dynamics we have identified two regimes: (i) $\gamma \ll \unit{\hopping}$: 
As the electrons are able to exchange energy with an environment the effective interaction made up from the combined electron--phonon and electron-electron interaction determines how fast the system relaxes.
Larger (effective) interactions lead to faster relaxation.
In the second regime (ii) $\gamma > \unit{\hopping}$, polaron formation leads to less mobile particles, thus a larger effective disorder strength is seen, and the relaxation slows down. 

\section{Conclusion}
\label{sec:conclusions}

In this work we have introduced two time-dependent methods that combine accurate numerical many-body method capable of accounting for strong, direct electronic interactions, with multitrajectory Ehrenfest dynamics.
These two methods are Lanczos-time propagation and TEBD, the latter based on matrix-product states. We primarily benchmark our methods in the case of the one-dimensional Holstein model by computing electronic observables. Both methods agree with each other and with standard MTE implementations for systems without direct electronic interactions.

The advantage of Lanczos are the long time scales that can be simulated while TEBD can access larger system sizes. Lanczos could readily be applied to small two-dimensional clusters as well. While not studied in this work, we envision the main application of TEBD-MTE (and extensions) in the calculation of dynamical properties in equilibrium and at finite temperatures. 

As a concrete application, we studied the decay of charge-density wave-order at half filling in the presence of repulsive electronic interactions and in the presence of quenched onsite disorder, with a classical phonon environment. 
For both noninteracting and interacting systems, the electron--phonon coupling leads to delocalization, witnessed by a decay of the charge-density-wave order parameter. In the simulated time window, this decay follows a power-law and is faster in the interacting case for small $\gamma/\unit{\hopping}$, as expected. 
The system shows subdiffusive dynamics, both in the single-electron case and for finite densities, at least on the simulated time scales.

There are several interesting extensions and open questions to our work. 
On the technical side, the application of TEBD-MTE to dynamical quantities --- e.g., spectral functions, and conductivities --- seems promising. Moreover, TEBD can be replaced by other time-propagating schemes for matrix-product states~\cite{Paeckel2019}. 
In the long run, a combination of matrix-product state methods and coupled-trajectory methods (such as, for instance multiconfigurational Ehrenfest~\cite{Shalashilin2009,Shalashilin2010}) may be lead to much more reliable results than MTE.

As for the physics of interacting electrons in the presence of \emph{disorder}, few numerical studies have been conducted that include phonons as a generic relaxation channel in a solid-state environment while treating electronic correlations at a \emph{finite} filling adequately \cite{Bronold2002,Bronold2004,Berciu2010}.
Many interesting questions present themselves, such as a full characterization of transport \cite{Miladic2025}, establishing a connection to variable-range hopping, and the extension to acoustic phonons, to name a few.\\

Simulation codes and the data shown in the figures is publicly available on Zenodo~\cite{this_zenodo}.\\

\textit{Note added:} During the final stages
of preparing this manuscript, we became aware
of a study of CDW dynamics in Holstein chains
using a classical approximation to the phonon sector \cite{Jang2025}.

\begin{acknowledgments}
    We acknowledge useful discussions with P. Bl\"ochl, M. Hopjan, A. Kelly, P. Sollich, M. ten Brink, D. Tapias, and J. Wang. This work was funded by the Deutsche Forschungsgemeinschaft (DFG, German Research Foundation) – 217133147, 499180199, 436382789, 493420525; via CRC 1073 (project B09), FOR 5522 and large-equipment grants (GOEGrid cluster).
    This research was supported in part by the National Science Foundation under Grant No. NSF PHY-1748958.
\end{acknowledgments}

\appendix

\section{Behavior of single trajectories in MTE}
\label{sec:appendix_chaos}

\begin{figure}[ht]
    \centering
    \includegraphics[width=\linewidth]{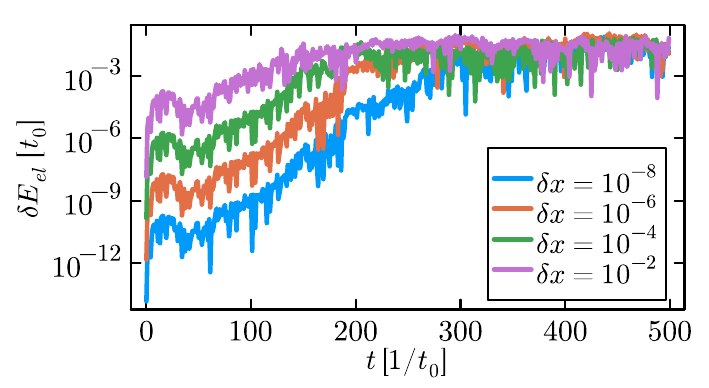}
    \caption{
        Dynamics in a single trajectory: Decay of the CDW in the non-interacting system (${L=100}$, ${V=0}$, ${\gamma = \qty{0.5}{\hopping}}$, ${W=\qty{8}{\hopping}}$, ${\omega_0=\qty{0.1}{\hopping}}$).
        We compare trajectories with slightly different initial conditions: 
        The trajectories only vary in their initial condition at site $\ell=50$ by \changes{$x_\ell(t=0) = x_\ell(t=0) + \delta x$}. 
        The data sets show the resulting difference in the electronic energy $\delta E_\mathrm{el} = |E_\mathrm{el}(0) - E_\mathrm{el}(\delta x) | $.
        The large increase of $\delta E_\mathrm{el}$ is seen over a wide range of investigated parameter regimes. 
    }
    \label{fig:appendix_sensitive_trajectories}
\end{figure}

Single trajectories under MTE dynamics may undergo chaotic dynamics.
It turns out that single trajectories show a \emph{sensitivity to initial conditions} as one would expect from a \emph{classically} chaotic system constituent of only classical degrees of freedom.
While the Ehrenfest dynamics make up a mixed quantum--classical system, the quantum subsystem acts like a time-dependent driving of the harmonic oscillators in the classical part of the system.
Indeed, driving a classical system time-dependently is one of the simplest ways to create sensitivity to initial conditions \cite{Schuster1995}.

This is reflected in the dependence of trajectories on initial conditions and numerical control parameters. As \cref{fig:convergence_lanczos_timestep,fig:convergence_tebd_timestep}
shows, the electronic energy develops a large deviation between originally
nearby trajectories, independent of the initial deviation in the coordinate at a single site. Moreover, we see that the initial state dependence of the classical system also leads to large deviations in the
electronic observables, see \cref{fig:appendix_sensitive_trajectories} where we have investigated the difference of the electronic energy $\delta E_\mathrm{el} = |E_\mathrm{el}(0) - E_\mathrm{el}(\delta x)|$ between slightly perturbed trajectories.
Here, $\delta x$ is a small perturbation acting on the harmonic oscillator position \vanishchanges{coordinate} $x_\ell$ at site $\ell = 50$ and time $t=0$ \vanishchanges{like so}, \changes{changing the initial condition to $x_\ell(t=0) = x_\ell(t=0) + \delta x$.} 
We observe that already a small perturbation $\delta x = 10^{-8}$ is enough to have large effects on the dynamics of the system, as would be expected in a classically chaotic system.

Fortunately, the ensemble average over many trajectories is not affected by this error proliferation.
As expected, only the ensemble average carries actual physical meaning.

\section{Convergence of TEBD-MTE for large systems}

\begin{figure}[ht]
    \centering
    \includegraphics{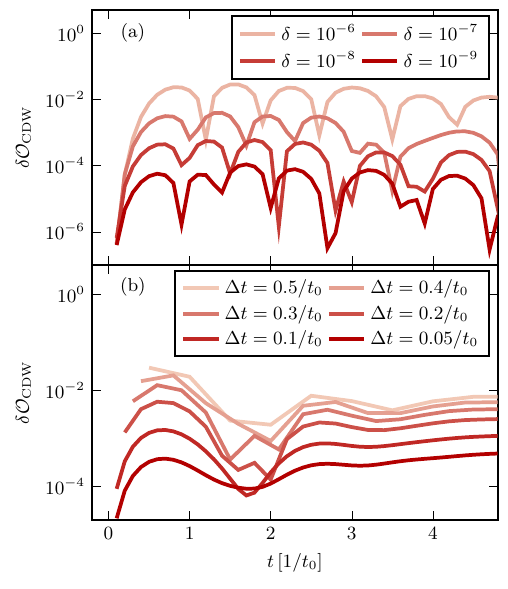}
    {\phantomsubcaption\label{fig:appendix_systemsize_cutoff}}
    {\phantomsubcaption\label{fig:appendix_systemsize_timestep}}
    \caption{ 
        Dependence of the TEBD-MTE method on the discarded weight $\delta$ and the time step $\Delta t$ in an interacting system ($L=28$, $V=\qty{2}{\hopping}$, $\omega_0=\qty{0.1}{\hopping}$, $\gamma=\qty{0.4}{\hopping}$, $W=0$).
        We plot the deviation $\delta \mathcal{O}_\mathrm{CDW}$ when \subref{fig:appendix_systemsize_cutoff} varying $\delta$ at a fixed time step $\Delta t = \qty{0.01}{\per\hopping}$, comparing with a reference set using $\delta = 10^{-10}$ and \subref{fig:appendix_systemsize_timestep} varying $\Delta t$ at fixed cutoff $\delta = 10^{-10}$, comparing with reference set using $\Delta t = \qty{0.01}{\per\hopping}$.
        The results are averaged over $N_\mathrm{traj} = 4000$ trajectories.
       }
    \label{fig:appendix_systemsize}
\end{figure}

\label{sec:L28}
In the main text, we present the results of the TEBD-MTE method for small systems. Because of the small system size of $L=14$, the bond dimension of the MPS reaches the maximum physical value. 
Here we consider a larger system of $L=28$. 
In this case, we can only time evolve the system for a shorter time before it reaches a given maximum bond dimension of the MPS considered in the simulation. 
We study the convergence with the time step and the truncation in the singular values as before (\cref{sec:DMRG-mte}) with a cut-off bond dimension of $\chi_{cut}=1000$. 

We plot the same quantities ($\delta O_{\mathrm{CDW}}$) that we plot in the \cref{fig:appendix_systemsize} for $\Delta t^* = \qty{0.01}{\per\hopping}$ and $\delta^* = 10^{-10}$ in \cref{fig:appendix_systemsize_timestep} and \cref{fig:appendix_systemsize_cutoff}, respectively. 
We can see that the convergence improves with decreasing $\Delta t$ and $\delta$. 
The maximum bond dimension $\chi_{\mathrm{max}}-\chi_{\mathrm{max}}(\delta)$ needed for a given discarded weight grows at different rates for different $\delta$. 
At time $t=\qty{4.4}{\per\hopping}$, $\chi_{\mathrm{max}}$ reaches $122,~283,~461,~664$ and $903$ for $\delta = 10^{-6},~10^{-7},~10^{-8},~10^{-9},$ and $~10^{-10}$, respectively. Moreover, we observe that the convergence is improved (smaller $\delta O_{\mathrm{CDW}}$) for this case compared to the smaller system considered in \cref{sec:DMRG-mte}. 
This may be because of self-averaging of the disorder potential in larger systems. 



\bibliography{references}

@article{Vidal2004,
  title = {Efficient Simulation of One-Dimensional Quantum Many-Body Systems},
  author = {Vidal, Guifr\'e},
  journal = {Phys. Rev. Lett.},
  volume = {93},
  issue = {4},
  pages = {040502},
  numpages = {4},
  year = {2004},
  month = {Jul},
  publisher = {American Physical Society},
  doi = {10.1103/PhysRevLett.93.040502},
  url = {https://link.aps.org/doi/10.1103/PhysRevLett.93.040502}
}

@article{Jang2025,
  author = {Ho Jang and Yang Yang and Gia-Wei Chern},
  title  = {Anomalous coarsening and nonlinear diffusion of kinks in an one-dimensional quasi-classical {H}olstein model},
  pages  = {arXiv:2512.07744},
  doi = {10.48550/arXiv.2512.07744},
  journal = {arXiv preprint},
  year = {2025}
}

@article{Paprotzki2023,
  title = {Quench dynamics in higher-dimensional {H}olstein models: {I}nsights from truncated {W}igner approaches},
  author = {Paprotzki, Eva and Osterkorn, Alexander and Mishra, Vibhu and Kehrein, Stefan},
  journal = {Phys. Rev. B},
  volume = {109},
  issue = {17},
  pages = {174303},
  numpages = {15},
  year = {2024},
  month = {May},
  publisher = {American Physical Society},
  doi = {10.1103/PhysRevB.109.174303},
  url = {https://link.aps.org/doi/10.1103/PhysRevB.109.174303}
}

@article{Tozer2014,
  title = {Localization of large polarons in the disordered {H}olstein model},
  author = {Tozer, Oliver Robert and Barford, William},
  journal = {Phys. Rev. B},
  volume = {89},
  issue = {15},
  pages = {155434},
  numpages = {7},
  year = {2014},
  month = {Apr},
  publisher = {American Physical Society},
  doi = {10.1103/PhysRevB.89.155434},
  url = {https://link.aps.org/doi/10.1103/PhysRevB.89.155434}
}

@article{Berciu2012,
  title = {Perturbational study of the lifetime of a {H}olstein polaron in the presence of weak disorder},
  author = {Ebrahimnejad, Hadi and Berciu, Mona},
  journal = {Phys. Rev. B},
  volume = {86},
  issue = {20},
  pages = {205109},
  numpages = {9},
  year = {2012},
  month = {Nov},
  publisher = {American Physical Society},
  doi = {10.1103/PhysRevB.86.205109},
  url = {https://link.aps.org/doi/10.1103/PhysRevB.86.205109}
}

@article{Berciu2010,
doi = {10.1209/0295-5075/89/37007},
url = {https://dx.doi.org/10.1209/0295-5075/89/37007},
year = {2010},
month = {feb},
publisher = {},
volume = {89},
number = {3},
pages = {37007},
author = {M. Berciu and A. S. Mishchenko and N. Nagaosa},
title = {{H}olstein polaron in the presence of disorder},
journal = {Europhysics Letters},
}

@article{Das2008,
doi = {10.1088/0953-8984/20/34/345222},
url = {https://dx.doi.org/10.1088/0953-8984/20/34/345222},
year = {2008},
month = {aug},
publisher = {},
volume = {20},
number = {34},
pages = {345222},
author = {A N Das and S Sil},
title = {Thermodynamic properties of {H}olstein polarons and the effects of disorder},
journal = {Journal of Physics: Condensed Matter},
}

@article{Bronold2002,
  title = {{A}nderson localization of polaron states},
  author = {Bronold, Franz X. and Fehske, Holger},
  journal = {Phys. Rev. B},
  volume = {66},
  issue = {7},
  pages = {073102},
  numpages = {4},
  year = {2002},
  month = {Aug},
  publisher = {American Physical Society},
  doi = {10.1103/PhysRevB.66.073102},
  url = {https://link.aps.org/doi/10.1103/PhysRevB.66.073102}
}

@article{Schoenle2021,
  title = {Eigenstate thermalization hypothesis through the lens of autocorrelation functions},
  author = {Sch\"onle, Christoph and Jansen, David and Heidrich-Meisner, Fabian and Vidmar, Lev},
  journal = {Phys. Rev. B},
  volume = {103},
  issue = {23},
  pages = {235137},
  numpages = {19},
  year = {2021},
  month = {Jun},
  publisher = {American Physical Society},
  doi = {10.1103/PhysRevB.103.235137},
  url = {https://link.aps.org/doi/10.1103/PhysRevB.103.235137}
}

@article{Fleishman1978,
  title = {Elementary Excitations in the {F}ermi Glass},
  author = {Fleishman, L. and Licciardello, D. C. and Anderson, P. W.},
  journal = {Phys. Rev. Lett.},
  volume = {40},
  issue = {20},
  pages = {1340--1343},
  numpages = {0},
  year = {1978},
  month = {May},
  publisher = {American Physical Society},
  doi = {10.1103/PhysRevLett.40.1340},
  url = {https://link.aps.org/doi/10.1103/PhysRevLett.40.1340}
}

@article{Pastor2022,
	author = {Pastor, Ernest and Sachs, Michael and Selim, Shababa and Durrant, James R. and Bakulin, Artem A. and Walsh, Aron},
	date = {2022/07/01},
	doi = {10.1038/s41578-022-00433-0},
	id = {Pastor2022},
	isbn = {2058-8437},
	journal = {Nature Reviews Materials},
	number = {7},
	pages = {503--521},
	title = {Electronic defects in metal oxide photocatalysts},
	url = {https://doi.org/10.1038/s41578-022-00433-0},
	volume = {7},
	year = {2022}}

@article{Schreiber2015,
    Author = {M. Schreiber and S. S. Hodgman and P. Bordia and H. P. L\"uschen and M. H. Fischer and R. Vosk and E. Altman and U. Schneider and I. Bloch},
        Journal = {Science},
        Pages = {842},
        Title = {Observation of many-body localization of interacting fermions in a quasi-random optical lattice},
        Volume = {349},
        Year = {2015},
        doi = {DOI: 10.1126/science.aaa74}
}

@article{Bronold2004,
    author = {F. X. Bronold and A. Alvermann and H. Fehske},
    title = {{A}nderson localization in strongly coupled disordered electron–phonon systems},
    journal = {Philosophical Magazine},
    volume = {84},
    number = {7},
    pages = {673--704},
    year = {2004},
    publisher = {Taylor \& Francis},
    doi = {10.1080/14786430310001624884}
}

@article{Bai2018,
    author = {Bai, Xin and Qiu, Jing and Wang, Linjun},
    title = {An efficient solution to the decoherence enhanced trivial crossing problem in surface hopping},
    journal = {The Journal of Chemical Physics},
    volume = {148},
    number = {10},
    pages = {104106},
    year = {2018},
    month = {03},
    doi = {https://doi.org/10.1063/1.5020693}
}

@Article{Carof2019,
author ="Carof, Antoine and Giannini, Samuele and Blumberger, Jochen",
title  ={How to calculate charge mobility in molecular materials from surface hopping non-adiabatic molecular dynamics – beyond the hopping/band paradigm},
journal  ="Phys. Chem. Chem. Phys.",
year  ="2019",
volume  ="21",
issue  ="48",
pages  ="26368-26386",
publisher  ="The Royal Society of Chemistry",
doi  ="10.1039/C9CP04770K",
url  ="http://dx.doi.org/10.1039/C9CP04770K",
}

@article{Jansen2023,
  title = {Thermal and optical conductivity in the {H}olstein model at half filling and finite temperature in the {L}uttinger-liquid and charge-density-wave regime},
  author = {Jansen, David and Heidrich-Meisner, Fabian},
  journal = {Phys. Rev. B},
  volume = {108},
  issue = {8},
  pages = {L081114},
  numpages = {7},
  year = {2023},
  month = {Aug},
  publisher = {American Physical Society},
  doi = {10.1103/PhysRevB.108.L081114},
  url = {https://link.aps.org/doi/10.1103/PhysRevB.108.L081114}
}

@article{Jansen2022,
  title = {Finite-temperature optical conductivity with density-matrix renormalization group methods for the {H}olstein polaron and bipolaron with dispersive phonons},
  author = {Jansen, David and Bon\ifmmode \check{c}\else \v{c}\fi{}a, Janez and Heidrich-Meisner, Fabian},
  journal = {Phys. Rev. B},
  volume = {106},
  issue = {15},
  pages = {155129},
  numpages = {21},
  year = {2022},
  month = {Oct},
  publisher = {American Physical Society},
  doi = {10.1103/PhysRevB.106.155129},
  url = {https://link.aps.org/doi/10.1103/PhysRevB.106.155129}
}

@article{Jansen2020,
  title = {Finite-temperature density-matrix renormalization group method for electron--phonon systems: Thermodynamics and {Ho}lstein-polaron spectral functions},
  author = {Jansen, David and Bon\ifmmode \check{c}\else \v{c}\fi{}a, Janez and Heidrich-Meisner, Fabian},
  journal = {Phys. Rev. B},
  volume = {102},
  issue = {16},
  pages = {165155},
  numpages = {17},
  year = {2020},
  month = {Oct},
  publisher = {American Physical Society},
  doi = {10.1103/PhysRevB.102.165155},
  url = {https://link.aps.org/doi/10.1103/PhysRevB.102.165155}
}

@article{Hashimoto2017,
  title = {Photoinduced charge-order melting dynamics in a one-dimensional interacting {H}olstein model},
  author = {Hashimoto, Hiroshi and Ishihara, Sumio},
  journal = {Phys. Rev. B},
  volume = {96},
  issue = {3},
  pages = {035154},
  numpages = {11},
  year = {2017},
  month = {Jul},
  publisher = {American Physical Society},
  doi = {10.1103/PhysRevB.96.035154},
  url = {https://link.aps.org/doi/10.1103/PhysRevB.96.035154}
}

@article{Jeckelmann1998,
  title = {Density-matrix renormalization-group study of the polaron problem in the {H}olstein model},
  author = {Jeckelmann, Eric and White, Steven R.},
  journal = {Phys. Rev. B},
  volume = {57},
  issue = {11},
  pages = {6376--6385},
  numpages = {0},
  year = {1998},
  month = {Mar},
  publisher = {American Physical Society},
  doi = {10.1103/PhysRevB.57.6376},
  url = {https://link.aps.org/doi/10.1103/PhysRevB.57.6376}
}

@article{Stolpp2020,
  title = {Charge-density-wave melting in the one-dimensional {H}olstein model},
  author = {Stolpp, Jan and Herbrych, Jacek and Dorfner, Florian and Dagotto, Elbio and Heidrich-Meisner, Fabian},
  journal = {Phys. Rev. B},
  volume = {101},
  issue = {3},
  pages = {035134},
  numpages = {18},
  year = {2020},
  month = {Jan},
  publisher = {American Physical Society},
  doi = {10.1103/PhysRevB.101.035134},
  url = {https://link.aps.org/doi/10.1103/PhysRevB.101.035134}
}

@article{Shalashilin2010,
    author = {Shalashilin, Dmitrii V.},
    title = {Nonadiabatic dynamics with the help of multiconfigurational Ehrenfest method: Improved theory and fully quantum 24D simulation of pyrazine},
    journal = {The Journal of Chemical Physics},
    volume = {132},
    number = {24},
    pages = {244111},
    year = {2010},
    month = {06},
    issn = {0021-9606},
    doi = {10.1063/1.3442747}
}

@article{Shalashilin2009,
    author = {Shalashilin, Dmitrii V.},
    title = "{Quantum mechanics with the basis set guided by {E}hrenfest trajectories: {T}heory and application to spin-boson model}",
    journal = {The Journal of Chemical Physics},
    volume = {130},
    number = {24},
    pages = {244101},
    year = {2009},
    month = {06},
    issn = {0021-9606},
    doi = {10.1063/1.3153302}
}

@article{Subotnik2016,
author = {Subotnik, Joseph E. and Jain, Amber and Landry, Brian and Petit, Andrew and Ouyang, Wenjun and Bellonzi, Nicole},
title = {Understanding the Surface Hopping View of Electronic Transitions and Decoherence},
journal = {Annual Review of Physical Chemistry},
volume = {67},
number = {1},
pages = {387-417},
year = {2016},
doi = {10.1146/annurev-physchem-040215-112245},
}

@article{Polkovnikov2010,
title = {Phase space representation of quantum dynamics},
journal = {Annals of Physics},
volume = {325},
number = {8},
pages = {1790-1852},
year = {2010},
issn = {0003-4916},
doi = {https://doi.org/10.1016/j.aop.2010.02.006},
url = {https://www.sciencedirect.com/science/article/pii/S0003491610000382},
author = {Anatoli Polkovnikov},
}

@article{Zhang1998,
  title = {Density Matrix Approach to Local {H}ilbert Space Reduction},
  author = {Zhang, Chunli and Jeckelmann, Eric and White, Steven R.},
  journal = {Phys. Rev. Lett.},
  volume = {80},
  issue = {12},
  pages = {2661--2664},
  numpages = {0},
  year = {1998},
  month = {Mar},
  publisher = {American Physical Society},
  doi = {10.1103/PhysRevLett.80.2661},
  url = {https://link.aps.org/doi/10.1103/PhysRevLett.80.2661}
}

@article{Weber2021,
  title = {Real-time evolution of static electron--phonon models in time-dependent electric fields},
  author = {Weber, Manuel and Freericks, James K.},
  journal = {Phys. Rev. E},
  volume = {105},
  issue = {2},
  pages = {025301},
  numpages = {10},
  year = {2022},
  month = {Feb},
  publisher = {American Physical Society},
  doi = {10.1103/PhysRevE.105.025301},
  url = {https://link.aps.org/doi/10.1103/PhysRevE.105.025301}
}

@article{Cohen-Stead2023,
title={{SmoQyDQMC.jl: A flexible implementation of determinant quantum {M}onte {C}arlo for {H}ubbard and electron--phonon interactions}},
	author={Benjamin Cohen-Stead and Sohan Malkaruge Costa and James Neuhaus and Andy Tanjaroon Ly and Yutan Zhang and Richard Scalettar and Kipton Barros and Steven Johnston},
	journal={SciPost Phys. Codebases},
	pages={29},
	year={2024},
	publisher={SciPost},
	doi={10.21468/SciPostPhysCodeb.29},
	url={https://scipost.org/10.21468/SciPostPhysCodeb.29},
}

@article{Kloss2019,
  title = {Multiset Matrix Product State Calculations Reveal Mobile {F}ranck-{C}ondon Excitations Under Strong {H}olstein-Type Coupling},
  author = {Kloss, Benedikt and Reichman, David R. and Tempelaar, Roel},
  journal = {Phys. Rev. Lett.},
  volume = {123},
  issue = {12},
  pages = {126601},
  numpages = {6},
  year = {2019},
  month = {Sep},
  publisher = {American Physical Society},
  doi = {10.1103/PhysRevLett.123.126601},
  url = {https://link.aps.org/doi/10.1103/PhysRevLett.123.126601}
}

@article{Stolpp2021,
title = {Comparative study of state-of-the-art matrix-product-state methods for lattice models with large local {H}ilbert spaces without {$U(1)$} symmetry},
journal = {Computer Physics Communications},
volume = {269},
pages = {108106},
year = {2021},
issn = {0010-4655},
doi = {https://doi.org/10.1016/j.cpc.2021.108106},
url = {https://www.sciencedirect.com/science/article/pii/S0010465521002186},
author = {Jan Stolpp and Thomas Köhler and Salvatore R. Manmana and Eric Jeckelmann and Fabian Heidrich-Meisner and Sebastian Paeckel},
}

@article{Guo2011,
  title = {Critical and Strong-Coupling Phases in One- and Two-Bath Spin-Boson Models},
  author = {Guo, Cheng and Weichselbaum, Andreas and von Delft, Jan and Vojta, Matthias},
  journal = {Phys. Rev. Lett.},
  volume = {108},
  issue = {16},
  pages = {160401},
  numpages = {5},
  year = {2012},
  month = {Apr},
  publisher = {American Physical Society},
  doi = {10.1103/PhysRevLett.108.160401},
  url = {https://link.aps.org/doi/10.1103/PhysRevLett.108.160401}
}

@Article{Brockt2015,
  Title                    = {Matrix-product-state method with a dynamical local basis optimization for bosonic systems out of equilibrium},
  Author                   = {Brockt, C. and Dorfner, F. and Vidmar, L. and Heidrich-Meisner, F. and Jeckelmann, E.},
  Journal                  = {Phys. Rev. B},
  Year                     = 2015,
Month                    = {Dec},
  Pages                    = 241106,
  Volume                   = 92,
Doi                      = {10.1103/PhysRevB.92.241106},
  Issue                    = 24,
  Numpages                 = 5,
  Publisher                = {American Physical Society},
  Url                      = {https://link.aps.org/doi/10.1103/PhysRevB.92.241106}
}

@article{Picano2023,
  title = {Stochastic semiclassical theory for nonequilibrium electron--phonon coupled systems},
  author = {Picano, Antonio and Grandi, Francesco and Werner, Philipp and Eckstein, Martin},
  journal = {Phys. Rev. B},
  volume = {108},
  issue = {3},
  pages = {035115},
  numpages = {16},
  year = {2023},
  month = {Jul},
  publisher = {American Physical Society},
  doi = {10.1103/PhysRevB.108.035115},
  url = {https://link.aps.org/doi/10.1103/PhysRevB.108.035115}
}

@Article{Dorfner2015,
  Title                    = {{Real-time decay of a highly excited charge carrier in the one-dimensional {H}olstein model}},
  Author                   = {Dorfner, F. and Vidmar, L. and Brockt, C. and Jeckelmann, E. and Heidrich-Meisner, F.},
  Journal                  = {Phys. Rev. B},
  Year                     = {2015},

  Month                    = {Mar},
  Pages                    = {104302},
  Volume                   = {91},

  Doi                      = {10.1103/PhysRevB.91.104302},
  Issue                    = {10},
  Numpages                 = {22},
  Publisher                = {American Physical Society}
}

@Article{Bonca1999,
  Title                    = {{{H}olstein polaron}},
  Author                   = {Bon\v{c}a, J. and Trugman, S. A. and Batisti\'{c}, I.},
  Journal                  = {Phys. Rev. B},
  Year                     = {1999},

  Month                    = {Jul},
  Pages                    = {1633--1642},
  Volume                   = {60},
  Doi                      = {10.1103/PhysRevB.60.1633},
  Issue                    = {3},
  Numpages                 = {0},
  Publisher                = {American Physical Society},
  Url                      = {http://link.aps.org/doi/10.1103/PhysRevB.60.1633}
}

@article{Bloch2022,
author = {Jacqueline Bloch and Andrea Cavalleri and Victor Galitski and Mohammad Hafezi and Angel Rubio},
journal = {Nature},
title  = {Strongly-Correlated Electron--Photon Systems },
year = {2022},
pages = {41},
volume = {606},
doi = {10.1038/s41586-022-04726-w}
}

@article{Murakami2015,
  title = {Interaction quench in the {H}olstein model: {T}hermalization crossover from electron- to phonon-dominated relaxation},
  author = {Murakami, Yuta and Werner, Philipp and Tsuji, Naoto and Aoki, Hideo},
  journal = {Phys. Rev. B},
  volume = {91},
  issue = {4},
  pages = {045128},
  numpages = {13},
  year = {2015},
  month = {Jan},
  publisher = {American Physical Society},
  doi = {10.1103/PhysRevB.91.045128},
  url = {https://link.aps.org/doi/10.1103/PhysRevB.91.045128}
}

@article{DiSante2017,
  title = {Disorder-Driven Metal-Insulator Transitions in Deformable Lattices},
  author = {Di Sante, Domenico and Fratini, Simone and Dobrosavljevi\ifmmode \acute{c}\else \'{c}\fi{}, Vladimir and Ciuchi, Sergio},
  journal = {Phys. Rev. Lett.},
  volume = {118},
  issue = {3},
  pages = {036602},
  numpages = {5},
  year = {2017},
  month = {Jan},
  publisher = {American Physical Society},
  doi = {10.1103/PhysRevLett.118.036602},
  url = {https://link.aps.org/doi/10.1103/PhysRevLett.118.036602}
}

@article{Manawadu2022,
author = {Manawadu, Dilhan and Valentine, Darren J. and Marcus, Max and Barford, William},
title = {Singlet Triplet-Pair Production and Possible Singlet-Fission in Carotenoids},
journal = {The Journal of Physical Chemistry Letters},
volume = {13},
number = {5},
pages = {1344-1349},
year = {2022},
doi = {10.1021/acs.jpclett.1c03812},   
}

@article{Manawadu2023,
author = {Manawadu, Dilhan and Valentine, Darren J. and Barford, William},
title = {Dynamical Simulations of Carotenoid Photoexcited States Using Density Matrix Renormalization Group Techniques},
journal = {The Journal of Physical Chemistry A},
volume = {127},
number = {16},
pages = {3714-3727},
year = {2023},
doi = {10.1021/acs.jpca.3c00988},
    note ={PMID: 37054397}
}

@article{Prelovsek2018,
  title = {Transient and persistent particle subdiffusion in a disordered chain coupled to bosons},
  author = {Prelov\ifmmode \check{s}\else \v{s}\fi{}ek, P. and Bon\ifmmode \check{c}\else \v{c}\fi{}a, J. and Mierzejewski, M.},
  journal = {Phys. Rev. B},
  volume = {98},
  issue = {12},
  pages = {125119},
  numpages = {6},
  year = {2018},
  month = {Sep},
  publisher = {American Physical Society},
  doi = {10.1103/PhysRevB.98.125119},
  url = {https://link.aps.org/doi/10.1103/PhysRevB.98.125119}
}

@article{Sierant2023,
  title = {Slow dynamics of a mobile impurity interacting with an {A}nderson insulator},
  author = {Sierant, Piotr and Chanda, Titas and Lewenstein, Maciej and Zakrzewski, Jakub},
  journal = {Phys. Rev. B},
  volume = {107},
  issue = {14},
  pages = {144201},
  numpages = {13},
  year = {2023},
  month = {Apr},
  publisher = {American Physical Society},
  doi = {10.1103/PhysRevB.107.144201},
  url = {https://link.aps.org/doi/10.1103/PhysRevB.107.144201}
}

@article{Mitric2025a,
  title = {Dynamical quantum typicality: {S}imple method for investigating transport properties applied to the {H}olstein model},
  author = {Mitri\ifmmode \acute{c}\else \'{c}\fi{}, Petar},
  journal = {Phys. Rev. B},
  volume = {111},
  issue = {19},
  pages = {195140},
  numpages = {10},
  year = {2025},
  month = {May},
  publisher = {American Physical Society},
  doi = {10.1103/PhysRevB.111.195140},
  url = {https://link.aps.org/doi/10.1103/PhysRevB.111.195140}
}

@article{Mitric2025,
  title = {Precursors to {A}nderson localization in the {H}olstein model: Quantum and quantum-classical solutions},
  author = {Mitri\ifmmode \acute{c}\else \'{c}\fi{}, P. and Dobrosavljevi\ifmmode \acute{c}\else \'{c}\fi{}, V. and Tanaskovi\ifmmode \acute{c}\else \'{c}\fi{}, D.},
  journal = {Phys. Rev. B},
  volume = {111},
  issue = {16},
  pages = {L161105},
  numpages = {5},
  year = {2025},
  month = {Apr},
  publisher = {American Physical Society},
  doi = {10.1103/PhysRevB.111.L161105},
  url = {https://link.aps.org/doi/10.1103/PhysRevB.111.L161105}
}

@article{Miladic2025,
  title = {Identification of the transport regimes of the one-dimensional {H}olstein model},
  author = {Miladi\ifmmode \acute{c}\else \'{c}\fi{}, Suzana and Vukmirovi\ifmmode \acute{c}\else \'{c}\fi{}, Nenad},
  journal = {Phys. Rev. B},
  volume = {112},
  issue = {5},
  pages = {054314},
  numpages = {14},
  year = {2025},
  month = {Aug},
  publisher = {American Physical Society},
  doi = {10.1103/t1g7-r95d},
  url = {https://link.aps.org/doi/10.1103/t1g7-r95d}
}

@article{Huang2024,
    author={Wayne Cheng-Wei Huang and Sai Mu and Gevin von Witte and Yanshuo Sophie Li and Felix Kurtz and Sheng-Hsiung Hung and Horng-Tay Jeng and Kai Rossnagel and Jan Gerrit Horstmann and Claus Ropers},
    title ={Ultrafast optical switching to a heterochiral charge-density wave state},
    year = {2024},
    pages = {arXiv:2405.20872},
    journal = {arXiv preprint},
    doi = {10.48550/arXiv.2405.20872},
    url = { 	
https://doi.org/10.48550/arXiv.2405.20872}
}

@article{Herbrych2012,
  title = {Spin hydrodynamics in the ${S}=\frac{1}{2}$ anisotropic {H}eisenberg chain},
  author = {Herbrych, J. and Steinigeweg, R. and Prelov\ifmmode \check{s}\else \v{s}\fi{}ek, P.},
  journal = {Phys. Rev. B},
  volume = {86},
  issue = {11},
  pages = {115106},
  numpages = {9},
  year = {2012},
  month = {Sep},
  publisher = {American Physical Society},
  doi = {10.1103/PhysRevB.86.115106},
  url = {https://link.aps.org/doi/10.1103/PhysRevB.86.115106}
}

@article{Steinigeweg2014,
  title = {Spin-Current Autocorrelations from Single Pure-State Propagation},
  author = {Steinigeweg, Robin and Gemmer, Jochen and Brenig, Wolfram},
  journal = {Phys. Rev. Lett.},
  volume = {112},
  issue = {12},
  pages = {120601},
  numpages = {5},
  year = {2014},
  month = {Mar},
  publisher = {American Physical Society},
  doi = {10.1103/PhysRevLett.112.120601},
  url = {https://link.aps.org/doi/10.1103/PhysRevLett.112.120601}
}

@article{Li2005,
    author = {Li, Xiaosong and Tully, John C. and Schlegel, H. Bernhard and Frisch, Michael J.},
    title = {Ab initio {E}hrenfest dynamics},
    journal = {The Journal of Chemical Physics},
    volume = {123},
    number = {8},
    pages = {084106},
    year = {2005},
    month = {08},
    issn = {0021-9606},
    doi = {10.1063/1.2008258},
    url = {https://doi.org/10.1063/1.2008258},
}

@article{Lively2023,

  title = {Revealing ultrafast phonon mediated inter-valley scattering through transient absorption and high harmonic spectroscopies},
  author = {Lively, Kevin and Sato, Shunsuke A. and Albareda, Guillermo and Rubio, Angel and Kelly, Aaron},
  journal = {Phys. Rev. Res.},
  volume = {6},
  issue = {1},
  pages = {013069},
  numpages = {18},
  year = {2024},
  month = {Jan},
  publisher = {American Physical Society},
  doi = {10.1103/PhysRevResearch.6.013069},
  url = {https://link.aps.org/doi/10.1103/PhysRevResearch.6.013069}
}

@article{Dagotto2001,
title = {Colossal magnetoresistant materials: {T}he key role of phase separation},
journal = {Physics Reports},
volume = {344},
number = {1},
pages = {1-153},
year = {2001},
issn = {0370-1573},
doi = {https://doi.org/10.1016/S0370-1573(00)00121-6},
url = {https://www.sciencedirect.com/science/article/pii/S0370157300001216},
author = {Elbio Dagotto and Takashi Hotta and Adriana Moreo},
}

@article{Tokura2006,
doi = {10.1088/0034-4885/69/3/R06},
url = {https://dx.doi.org/10.1088/0034-4885/69/3/R06},
year = {2006},
month = {feb},
publisher = {},
volume = {69},
number = {3},
pages = {797},
author = {Y Tokura},
title = {Critical features of colossal magnetoresistive manganites},
journal = {Reports on Progress in Physics},
}

@article{Rini2007,
	author = {Rini, Matteo and Tobey, Ra'anan and Dean, Nicky and Itatani, Jiro and Tomioka, Yasuhide and Tokura, Yoshinori and Schoenlein, Robert W. and Cavalleri, Andrea},
	date = {2007/09/01},
	doi = {10.1038/nature06119},
	id = {Rini2007},
	isbn = {1476-4687},
	journal = {Nature},
	number = {7158},
	pages = {72--74},
	title = {Control of the electronic phase of a manganite by mode-selective vibrational excitation},
	url = {https://doi.org/10.1038/nature06119},
	volume = {449},
	year = {2007}
}

@article{Tobey2008,
  title = {Ultrafast Electronic Phase Transition in {${\mathrm{La}}_{1/2}{\mathrm{Sr}}_{3/2}{\mathrm{MnO}}_{4}$} by Coherent Vibrational Excitation: Evidence for Nonthermal Melting of Orbital Order},
  author = {Tobey, R. I. and Prabhakaran, D. and Boothroyd, A. T. and Cavalleri, A.},
  journal = {Phys. Rev. Lett.},
  volume = {101},
  issue = {19},
  pages = {197404},
  numpages = {4},
  year = {2008},
  month = {Nov},
  publisher = {American Physical Society},
  doi = {10.1103/PhysRevLett.101.197404},
  url = {https://link.aps.org/doi/10.1103/PhysRevLett.101.197404}
}

@Article{Giannetti2016,
  Title                    = {Ultrafast optical spectroscopy of strongly correlated materials and high-temperature superconductors: a non-equilibrium approach},
  Author                   = {Claudio Giannetti and Massimo Capone and Daniele Fausti and Michele Fabrizio and Fulvio Parmigiani and Dragan Mihailovic},
  Journal                  = {Adv. Phys.},
  Year                     = {2016},
  Number                   = {2},
  Pages                    = {58-238},
  Volume                   = {65},
  
  Doi                      = {10.1080/00018732.2016.1194044},
  Publisher                = {Taylor \& Francis}, 
  Url                      = {https://doi.org/10.1080/00018732.2016.1194044} 
}

@article{TenBrink2022,
    author = {ten Brink, M. and Gräber, S. and Hopjan, M. and Jansen, D. and Stolpp, J. and Heidrich-Meisner, F. and Blöchl, P. E.},
    title = {Real-time non-adiabatic dynamics in the one-dimensional {H}olstein model: {T}rajectory-based vs exact methods},
    journal = {The Journal of Chemical Physics},
    volume = {156},
    number = {23},
    pages = {234109},
    year = {2022},
    month = {06},
    issn = {0021-9606},
    doi = {10.1063/5.0092063},
}

@article{Schollwoeck2011, 
title = {The density-matrix renormalization group in the age of matrix product states},
journal = "Ann. Phys. (N. Y.)",
volume = "326",
number = "1",
pages = "96 - 192",
year = "2011", 
issn = "0003-4916",
doi = "https://doi.org/10.1016/j.aop.2010.09.012",
url = "http://www.sciencedirect.com/science/article/pii/S0003491610001752",
author = {Ulrich Schollw\"ock},
}

@article{Goodvin2008,
  title = {Momentum average approximation for models with electron--phonon coupling dependent on the phonon momentum},
  author = {Goodvin, Glen L. and Berciu, Mona},
  journal = {Phys. Rev. B},
  volume = {78},
  issue = {23},
  pages = {235120},
  numpages = {13},
  year = {2008},
  month = {Dec},
  publisher = {American Physical Society},
  doi = {10.1103/PhysRevB.78.235120},
  url = {https://link.aps.org/doi/10.1103/PhysRevB.78.235120}
}

@article{Jankovic2022,
  title = {Spectral and thermodynamic properties of the {H}olstein polaron: {H}ierarchical equations  of motion approach},
  author = {Jankovi\ifmmode \acute{c}\else \'{c}\fi{}, Veljko and Vukmirovi\ifmmode \acute{c}\else \'{c}\fi{}, Nenad},
  journal = {Phys. Rev. B},
  volume = {105},
  issue = {5},
  pages = {054311},
  numpages = {22},
  year = {2022},
  month = {Feb},
  publisher = {American Physical Society},
  doi = {10.1103/PhysRevB.105.054311},
  url = {https://link.aps.org/doi/10.1103/PhysRevB.105.054311}
}

@article{White1993,
  title = {Density-matrix algorithms for quantum renormalization groups},
  author = {White, Steven R.},
  journal = {Phys. Rev. B},
  volume = {48},
  issue = {14},
  pages = {10345--10356},
  numpages = {0},
  year = {1993},
  month = {Oct},
  publisher = {American Physical Society},
  doi = {10.1103/PhysRevB.48.10345},
  url = {https://link.aps.org/doi/10.1103/PhysRevB.48.10345}
}

@article{White1992,
  title = {Density matrix formulation for quantum renormalization groups},
  author = {White, Steven R.},
  journal = {Phys. Rev. Lett.},
  volume = {69},
  issue = {19},
  pages = {2863--2866},
  numpages = {0},
  year = {1992},
  month = {Nov},
  publisher = {American Physical Society},
  doi = {10.1103/PhysRevLett.69.2863},
  url = {https://link.aps.org/doi/10.1103/PhysRevLett.69.2863}
}

@article{Paeckel2019,
title = {Time-evolution methods for matrix-product states},
journal = "Ann. Phys. (N. Y.)",
volume = "411",            
pages = "167998",
year = "2019",
issn = "0003-4916",
doi = "https://doi.org/10.1016/j.aop.2019.167998",
url = "http://www.sciencedirect.com/science/article/pii/S0003491619302532",
author = {Sebastian Paeckel and Thomas K\"ohler and Andreas Swoboda and Salvatore R. Manmana and Ulrich Schollw\"ock and Claudius Hubig},
}

@article{Sierant2023b,
  title = {Stability of many-body localization in {F}loquet systems},
  author = {Sierant, Piotr and Lewenstein, Maciej and Scardicchio, Antonello and Zakrzewski, Jakub},
  journal = {Phys. Rev. B},
  volume = {107},
  issue = {11},
  pages = {115132},
  numpages = {16},
  year = {2023},
  month = {Mar},
  publisher = {American Physical Society},
  doi = {10.1103/PhysRevB.107.115132},
  url = {https://link.aps.org/doi/10.1103/PhysRevB.107.115132}
}

@article{Manmana2005,
author={S. R. Manmana and R. Noack and A. Muramatsu},
title = {Time evolution of one-dimensional
Quantum Many Body Systems},
journal ={AIP Conf. Proc.}, 
volume ={789},
pages = {269},
year = {2005},
doi={ 	
https://doi.org/10.1063/1.208035}
}

@misc{Haegeman2024,
    author = {Haegeman, Jutho},
    doi = {10.5281/zenodo.10622234},
    month = mar,
    title = {{KrylovKit}},
    url = {https://github.com/Jutho/KrylovKit.jl},
    version = {0.6.1},
    year = {2024}
}

@article{itensor,
	title={{The ITensor Software Library for Tensor Network Calculations}},
	author={Matthew Fishman and Steven R. White and E. Miles Stoudenmire},
	journal={SciPost Phys. Codebases},
	pages={4},
	year={2022},
	publisher={SciPost},
	doi={10.21468/SciPostPhysCodeb.4},
	url={https://scipost.org/10.21468/SciPostPhysCodeb.4}
}

@BOOK{Schuster1995,
AUTHOR = {H. G. Schuster},
TITLE = {Deterministic Chaos: An Introduction},
PUBLISHER = {Wiley-VCH},
YEAR = {1995},
ADDRESS = {Philadelphia, PA},
ISBN = {3527290885} 
}

@article{Evers2008,
  author = {Evers, Ferdinand and Mirlin, Alexander D.},
  doi = {10.1103/revmodphys.80.1355},
  issue = {4},
  journal = {Reviews of Modern Physics},
  month = {10},
  pages = {1355--1417},
  publisher = {American Physical Society (APS)},
  title = {{A}nderson transitions},
  url = {http://dx.doi.org/10.1103/revmodphys.80.1355},
  volume = {80},
  year = {2008},
}

@article{Gogolin2016,
doi = {10.1088/0034-4885/79/5/056001},
url = {https://dx.doi.org/10.1088/0034-4885/79/5/056001},
year = {2016},
month = {apr},
publisher = {IOP Publishing},
volume = {79},
number = {5},
pages = {056001},
author = {Christian Gogolin and Jens Eisert},
title = {Equilibration, thermalisation, and the emergence of statistical mechanics in closed quantum systems},
journal = {Reports on Progress in Physics},
}

@article{Mori2018,
doi = {10.1088/1361-6455/aabcdf},
url = {https://dx.doi.org/10.1088/1361-6455/aabcdf},
year = {2018},
month = {may},
publisher = {IOP Publishing},
volume = {51},
number = {11},
pages = {112001},
author = {Takashi Mori and Tatsuhiko N Ikeda and Eriko Kaminishi and Masahito Ueda},
title = {Thermalization and prethermalization in isolated quantum systems: a theoretical overview},
journal = {Journal of Physics B: Atomic, Molecular and Optical Physics},
}

@article{Deutsch2018,
doi = {10.1088/1361-6633/aac9f1},
url = {https://dx.doi.org/10.1088/1361-6633/aac9f1},
year = {2018},
month = {jul},
publisher = {IOP Publishing},
volume = {81},
number = {8},
pages = {082001},
author = {Joshua M Deutsch},
title = {Eigenstate thermalization hypothesis},
journal = {Reports on Progress in Physics},
}

@article{Abanin2019,
  title = {Colloquium: {M}any-body localization, thermalization, and entanglement},
  author = {Abanin, Dmitry A. and Altman, Ehud and Bloch, Immanuel and Serbyn, Maksym},
  journal = {Rev. Mod. Phys.},
  volume = {91},
  issue = {2},
  pages = {021001},
  numpages = {26},
  year = {2019},
  month = {May},
  publisher = {American Physical Society},
  doi = {10.1103/RevModPhys.91.021001},
  url = {https://link.aps.org/doi/10.1103/RevModPhys.91.021001}
}

@article{dAlessio2016,
author = {Luca D'Alessio and Yariv Kafri and Anatoli Polkovnikov and Marcos Rigol},
title = {From quantum chaos and eigenstate thermalization to statistical mechanics and thermodynamics},
journal = {Adv. Phys.},
volume = {65},
pages = {239-362},
year  = {2016},
url={https://doi.org/10.1080/00018732.2016.1198134}
}

@article{Abanin2023,
  title = {Many-body localization proximity effect in a two-species bosonic {H}ubbard model},
  author = {Brighi, Pietro and Ljubotina, Marko and Abanin, Dmitry A. and Serbyn, Maksym},
  journal = {Phys. Rev. B},
  volume = {108},
  issue = {5},
  pages = {054201},
  numpages = {18},
  year = {2023},
  month = {Aug},
  publisher = {American Physical Society},
  doi = {10.1103/PhysRevB.108.054201}
}

@article{Bloch2019,
  title = {Many-Body Delocalization in the Presence of a Quantum Bath},
  author = {Rubio-Abadal, Antonio and Choi, Jae-yoon and Zeiher, Johannes and Hollerith, Simon and Rui, Jun and Bloch, Immanuel and Gross, Christian},
  journal = {Phys. Rev. X},
  volume = {9},
  issue = {4},
  pages = {041014},
  numpages = {6},
  year = {2019},
  month = {Oct},
  publisher = {American Physical Society},
  doi = {10.1103/PhysRevX.9.041014}
}

@article{Greiner2023,
	author = {L{\'e}onard, Julian and Kim, Sooshin and Rispoli, Matthew and Lukin, Alexander and Schittko, Robert and Kwan, Joyce and Demler, Eugene and Sels, Dries and Greiner, Markus},
	date = {2023/04/01},
	doi = {10.1038/s41567-022-01887-3},
	id = {L{\'e}onard2023},
	isbn = {1745-2481},
	journal = {Nature Physics},
	number = {4},
	pages = {481--485},
	title = {Probing the onset of quantum avalanches in a many-body localized system},
	volume = {19},
	year = {2023}
}

@article{Nandkishore2015,
  title = {Many-body localization proximity effect},
  author = {Nandkishore, Rahul},
  journal = {Phys. Rev. B},
  volume = {92},
  issue = {24},
  pages = {245141},
  numpages = {5},
  year = {2015},
  month = {Dec},
  publisher = {American Physical Society},
  doi = {10.1103/PhysRevB.92.245141}
}

@article{Sarang2017,
  title = {Noise-Induced Subdiffusion in Strongly Localized Quantum Systems},
  author = {Gopalakrishnan, Sarang and Islam, K. Ranjibul and Knap, Michael},
  journal = {Phys. Rev. Lett.},
  volume = {119},
  issue = {4},
  pages = {046601},
  numpages = {6},
  year = {2017},
  month = {Jul},
  publisher = {American Physical Society},
  doi = {10.1103/PhysRevLett.119.046601}
}

@article{Luitz2017a,
author = {Luitz, David J. and Lev, Yevgeny Bar},
title = {The ergodic side of the many-body localization transition},
journal = {Annalen der Physik},
volume = {529},
number = {7},
pages = {1600350},
doi = {https://doi.org/10.1002/andp.201600350},
year = {2017}
}

@article{Luitz2016a,
  title = {Extended slow dynamical regime close to the many-body localization transition},
  author = {Luitz, David J. and Laflorencie, Nicolas and Alet, Fabien},
  journal = {Phys. Rev. B},
  volume = {93},
  issue = {6},
  pages = {060201},
  numpages = {5},
  year = {2016},
  month = {Feb},
  publisher = {American Physical Society},
  doi = {10.1103/PhysRevB.93.060201},
  url = {https://link.aps.org/doi/10.1103/PhysRevB.93.060201}
}

@article{Luitz2017,
  title = {How a Small Quantum Bath Can Thermalize Long Localized Chains},
  author = {Luitz, David J. and Huveneers, Francois and De Roeck, Wojciech},
  journal = {Phys. Rev. Lett.},
  volume = {119},
  issue = {15},
  pages = {150602},
  numpages = {6},
  year = {2017},
  month = {Oct},
  publisher = {American Physical Society},
  doi = {10.1103/PhysRevLett.119.150602}
}

@article{Pollmann2020,
  title = {Entanglement dynamics of a many-body localized system coupled to a bath},
  author = {Wybo, Elisabeth and Knap, Michael and Pollmann, Frank},
  journal = {Phys. Rev. B},
  volume = {102},
  issue = {6},
  pages = {064304},
  numpages = {11},
  year = {2020},
  month = {Aug},
  publisher = {American Physical Society},
  doi = {10.1103/PhysRevB.102.064304}
}

@article{Abanin12021,
  title = {Nucleation of Ergodicity by a Single Mobile Impurity in Supercooled Insulators},
  author = {Krause, Ulrich and Pellegrin, Th\'eo and Brouwer, Piet W. and Abanin, Dmitry A. and Filippone, Michele},
  journal = {Phys. Rev. Lett.},
  volume = {126},
  issue = {3},
  pages = {030603},
  numpages = {6},
  year = {2021},
  month = {Jan},
  publisher = {American Physical Society},
  doi = {10.1103/PhysRevLett.126.030603}
}

@article{Katharine2017,
  title = {Many-body localization in the presence of a small bath},
  author = {Hyatt, Katharine and Garrison, James R. and Potter, Andrew C. and Bauer, Bela},
  journal = {Phys. Rev. B},
  volume = {95},
  issue = {3},
  pages = {035132},
  numpages = {8},
  year = {2017},
  month = {Jan},
  publisher = {American Physical Society},
  doi = {10.1103/PhysRevB.95.035132}
}

@article{HuseNandkishore2015,
  title = {Localized systems coupled to small baths: {F}rom {A}nderson to {Z}eno},
  author = {Huse, David A. and Nandkishore, Rahul and Pietracaprina, Francesca and Ros, Valentina and Scardicchio, Antonello},
  journal = {Phys. Rev. B},
  volume = {92},
  issue = {1},
  pages = {014203},
  numpages = {11},
  year = {2015},
  month = {Jul},
  publisher = {American Physical Society},
  doi = {10.1103/PhysRevB.92.014203}
}

@article{Bar2022,
	title={Logarithmic, noise-induced dynamics in the {A}nderson insulator},
	author={Talía L. M. Lezama and Yevgeny Bar Lev},
	journal={SciPost Phys.},
	volume={12},
	pages={174},
	year={2022},
	publisher={SciPost},
	doi={10.21468/SciPostPhys.12.5.174}
}

@article{Abanin2022,
  title = {Localization of a mobile impurity interacting with an {A}nderson insulator},
  author = {Brighi, Pietro and Michailidis, Alexios A. and Kirova, Kristina and Abanin, Dmitry A. and Serbyn, Maksym},
  journal = {Phys. Rev. B},
  volume = {105},
  issue = {22},
  pages = {224208},
  numpages = {18},
  year = {2022},
  month = {Jun},
  publisher = {American Physical Society},
  doi = {10.1103/PhysRevB.105.224208}
}

@article{SerbynAbanin2022,
  title = {Propagation of many-body localization in an {A}nderson insulator},
  author = {Brighi, Pietro and Michailidis, Alexios A. and Abanin, Dmitry A. and Serbyn, Maksym},
  journal = {Phys. Rev. B},
  volume = {105},
  issue = {22},
  pages = {L220203},
  numpages = {6},
  year = {2022},
  month = {Jun},
  publisher = {American Physical Society},
  doi = {10.1103/PhysRevB.105.L220203}
}

@article{Frohlich1954,
	author = {H. Fr{\"o}hlich},
	doi = {10.1080/00018735400101213},
	journal = {Advances in Physics},
	number = {11},
	pages = {325-361},
	publisher = {Taylor & Francis},
	title = {Electrons in lattice fields},
	volume = {3},
	year = {1954}
	}

@article{BCS1957,
  title = {Theory of Superconductivity},
  author = {Bardeen, J. and Cooper, L. N. and Schrieffer, J. R.},
  journal = {Phys. Rev.},
  volume = {108},
  issue = {5},
  pages = {1175--1204},
  numpages = {0},
  year = {1957},
  month = {Dec},
  publisher = {American Physical Society},
  doi = {10.1103/PhysRev.108.1175}
}

@article{Tully1998,
  author = {Tully, John C.},
  doi = {10.1039/a801824c},
  journal = {Faraday Discussions},
  pages = {407--419},
  publisher = {Royal Society of Chemistry (RSC)},
  title = {Mixed quantum\textendash{}classical dynamics},
  url = {http://dx.doi.org/10.1039/a801824c},
  volume = {110},
  year = {1998},
}

@article{Nielsen2001,
  author = {Nielsen, Steve and Kapral, Raymond and Ciccotti, Giovanni},
  doi = {10.1063/1.1400129},
  issue = {13},
  journal = {The Journal of Chemical Physics},
  month = {10},
  pages = {5805--5815},
  publisher = {AIP Publishing},
  title = {Statistical mechanics of quantum-classical systems},
  url = {http://dx.doi.org/10.1063/1.1400129},
  volume = {115},
  year = {2001},
}

@inbook{Grunwald2009,
  author = {Grunwald, Robbie and Kelly, Aaron and Kapral, Raymond},
  doi = {10.1007/978-3-642-02306-4\_12},
  isbn = {['9783642023057', '9783642023064']},
  journal = {Springer Series in Chemical Physics},
  month = {8},
  pages = {383--413},
  publisher = {Springer Berlin Heidelberg},
  title = {Quantum Dynamics in Almost Classical Environments},
  url = {http://dx.doi.org/10.1007/978-3-642-02306-4\_12},
  year = {2009},
}

@article{Wigner1932,
  author = {Wigner, E.},
  doi = {10.1103/physrev.40.749},
  issue = {5},
  journal = {Physical Review},
  month = {6},
  pages = {749--759},
  publisher = {American Physical Society (APS)},
  title = {On the Quantum Correction For Thermodynamic Equilibrium},
  url = {http://dx.doi.org/10.1103/physrev.40.749},
  volume = {40},
  year = {1932},
}

@article{Imre1967,
  author = {\.{I}mre, Kaya and \"{O}zizmir, Erc\"{u}ment and Rosenbaum, Marcos and Zweifel, P. F.},
  doi = {10.1063/1.1705323},
  issue = {5},
  journal = {Journal of Mathematical Physics},
  month = {5},
  pages = {1097--1108},
  publisher = {AIP Publishing},
  title = {{W}igner Method in Quantum Statistical Mechanics},
  url = {http://dx.doi.org/10.1063/1.1705323},
  volume = {8},
  year = {1967},
}

@article{Aleksandrov1981,
  author = {Aleksandrov, I. V.},
  doi = {10.1515/zna-1981-0819},
  issue = {8},
  journal = {Zeitschrift f\"ur Naturforschung A},
  month = {8},
  pages = {902--908},
  publisher = {Walter de Gruyter GmbH},
  title = {The Statistical Dynamics of a System Consisting of a Classical and a Quantum Subsystem},
  url = {http://dx.doi.org/10.1515/zna-1981-0819},
  volume = {36},
  year = {1981},
}

@article{Ehrenfest1927,
    author = {P. Ehrenfest},
    year = {1927},
    title = {{B}emerkung \"uber die angen\"aherte {G}\"ultigkeit der klassischen {M}echanik innerhalb der {Q}uantenmechanik},
    volume={45}, 
    pages = {455}, 
    journal = {Zeitschrift f\"ur Physik},
    doi = {https://doi.org/10.1039/a801824c}
}

@inbook{Kirrander2020,
    author = {Kirrander, Adam and Vacher, Morgane},
    publisher = {John Wiley \& Sons, Ltd},
    isbn = {9781119417774},
    title = {Ehrenfest Methods for Electron and Nuclear Dynamics},
    booktitle = {Quantum Chemistry and Dynamics of Excited States},
    chapter = {15},
    pages = {469-497},
    doi = {https://doi.org/10.1002/9781119417774.ch15},
    url = {https://onlinelibrary.wiley.com/doi/abs/10.1002/9781119417774.ch15},
    eprint = {https://onlinelibrary.wiley.com/doi/pdf/10.1002/9781119417774.ch15},
    year = {2020},
}

@article{Nibbering2005,
   author = "Nibbering, Erik T.J. and Fidder, Henk and Pines, Ehud",
   title = {Ultrafast chemistry: {U}sing Time-Resolved Vibrational Spectroscopy for Interrogation of Structural Dynamics},
   journal= "Annual Review of Physical Chemistry",
   year = "2005",
   volume = "56",
   number = "Volume 56, 2005",
   pages = "337-367",
   doi = "https://doi.org/10.1146/annurev.physchem.56.092503.141314",
   publisher = "Annual Reviews",
   issn = "1545-1593",
   type = "Journal Article",
}

@article{Parandekar2005,
  author = {Parandekar, Priya V. and Tully, John C.},
  doi = {10.1063/1.1856460},
  issue = {9},
  journal = {The Journal of Chemical Physics},
  month = {3},
  publisher = {AIP Publishing},
  title = {Mixed quantum-classical equilibrium},
  url = {http://dx.doi.org/10.1063/1.1856460},
  volume = {122},
  year = {2005},
  pages = {094102}
}

@article{Parandekar2006,
  author = {Parandekar, Priya V. and Tully, John C.},
  doi = {10.1021/ct050213k},
  issue = {2},
  journal = {Journal of Chemical Theory and Computation},
  month = {3},
  pages = {229--235},
  publisher = {American Chemical Society (ACS)},
  title = {Detailed Balance in Ehrenfest Mixed Quantum-Classical Dynamics},
  url = {http://dx.doi.org/10.1021/ct050213k},
  volume = {2},
  year = {2006},
}

@article{Kaeb2002,
  author = {K\"{a}b, G\"{u}nter},
  doi = {10.1103/physreve.66.046117},
  issue = {4},
  journal = {Phys. Rev. E},
  month = {10},
  publisher = {American Physical Society (APS)},
  title = {Mean field {E}hrenfest quantum/classical simulation of vibrational energy relaxation in a simple liquid},
  volume = {66},
  year = {2002},
 pages = {046117} 
}

@article{Park1986,
  author = {Jun Park, Tae and C. Light, J.},
  date = {1998-08-31},
  doi = {10.1063/1.451548},
  issn = {0021-9606},
  issue = {10},
  journal = {The Journal of Chemical Physics},
  month = {11},
  pages = {5870--5876},
  publisher = {American Institute of PhysicsAIP},
  title = {Unitary quantum time evolution by iterative {L}anczos reduction},
  url = {https://aip.scitation.org/doi/10.1063/1.451548},
  volume = {85},
  year = {1986},
}

@article{Kramer1993,
  author = {Kramer, B and MacKinnon, A},
  doi = {10.1088/0034-4885/56/12/001},
  issue = {12},
  journal = {Reports on Progress in Physics},
  month = {12},
  pages = {1469--1564},
  publisher = {IOP Publishing},
  title = {Localization: {T}heory and experiment},
  url = {http://dx.doi.org/10.1088/0034-4885/56/12/001},
  volume = {56},
  year = {1993},
}

@article{McLachlan2002,
  author = {I. McLachlan, Robert and Reinout W. Quispel, G.},
  doi = {10.1017/S0962492902000053},
  issn = {1474-0508},
  journal = {Acta Numerica},
  month = {1},
  pages = {341--434},
  publisher = {Cambridge University Press},
  title = {Splitting methods},
  url = {https://www.cambridge.org/core/journals/acta-numerica/article/splitting-methods/122F5736DAF3D88598989E68FE4D2EF2},
  volume = {11},
  year = {2002},
}

@article{Giustino2017,
  title = {Electron--phonon interactions from first principles},
  author = {Giustino, Feliciano},
  journal = {Rev. Mod. Phys.},
  volume = {89},
  issue = {1},
  pages = {015003},
  numpages = {63},
  year = {2017},
  month = {Feb},
  publisher = {American Physical Society},
  doi = {10.1103/RevModPhys.89.015003},
  url = {https://link.aps.org/doi/10.1103/RevModPhys.89.015003}
}

@dataset{this_zenodo,
    title={Hybrid quantum–classical matrix-product state and {L}anczos methods for electron–phonon systems with strong electronic correlations: {A}pplication to disordered systems coupled to {E}instein phonons},
    author={Menzler, Heiko Georg and Mondal, Suman and Heidrich-Meisner, Fabian},
    doi={10.5281/zenodo.17815721},
    month={Dec},
    year={2025}
}

@article{Koehler2021,
	title={{Efficient and flexible approach to simulate low-dimensional quantum lattice models with large local Hilbert spaces}},
	author={Thomas Köhler and Jan Stolpp and Sebastian Paeckel},
	journal={SciPost Phys.},
	volume={10},
	pages={058},
	year={2021},
	publisher={SciPost},
	doi={10.21468/SciPostPhys.10.3.058},
	url={https://scipost.org/10.21468/SciPostPhys.10.3.058},
}

\end{document}